\documentclass[aps,prd,email,showpacs,showkeys,preprintnumbers,amsmath,amssymb,nofootinbib,superscriptaddress,twocolumn]{revtex4}
\usepackage{slashed}
\usepackage[dvips]{graphicx}
\usepackage{pstricks,pst-coil}
\usepackage{pst-all}
\RequirePackage{flushend}
\RequirePackage[colorlinks,citecolor=blue,urlcolor=blue,linkcolor=blue]{hyperref}
\usepackage{color}
\usepackage{latexsym}
\usepackage{amsmath}
\usepackage{amssymb}
\usepackage{eufrak}
\usepackage{euscript}
\usepackage{graphics}
\usepackage{picture}
\def\[{\left\lbrack}
\def\]{\right\rbrack}

\def\({\left(}
\def\){\right)}

\newcommand{\bbe}{\begin{equation}}
\newcommand{\eee}{\end{equation}}
\newcommand{\eaa}{\end{eqnarray}}
\newcommand{\baa}{\begin{eqnarray}}

\def\ni{\noindent}

\def\uma{\rm 1\!\!\hskip 1 pt l}

\begin{document}


\title{\Large{The protophobic $X$-boson coupled to quantum electrodynamics}}

\author{Mario J. Neves}
\email{mariojr@ufrrj.br}
\affiliation{Grupo de F\'isica Te\'orica e
F\'isica Matem\'atica, Departamento de F\'{\i}sica,
Universidade Federal Rural do Rio de Janeiro,
BR 465-07, 23890-971, Serop\'edica, RJ, Brazil}

\author{Everton M. C. Abreu}
\email{evertonabreu@ufrrj.br}
\affiliation{Grupo de F\'isica Te\'orica e
F\'isica Matem\'atica, Departamento de F\'{\i}sica,
Universidade Federal Rural do Rio de Janeiro,
BR 465-07, 23890-971, Serop\'edica, RJ, Brazil}
\affiliation{Departamento de
F\'isica, Universidade Federal de Juiz de Fora, 36036-330, Juiz de
Fora, MG, Brazil}

\date{\today}



\begin{abstract}
\noindent The possible origin of an $X$-boson having a mass value around  $17 \, \mbox{MeV}$ had motivated us
to investigate its interaction with leptons of QED.
This new hypothetical particle can possibly be a candidate to describe the so called fifth interaction, in a new physics scenario
beyond the Standard Model. The model of the $X$-boson interacting with QED is based on a $U(1) \times U'(1)$ symmetry,
where the group $U'(1)$ is attached to the $X$-boson, with a kinetic mixing with the photon.
The Higgs sector was revisited to generate the mass for the new boson.   Thus, the mass of $17 \, \mbox{MeV}$ fixes a vacuum
expected value scale. Thereby, we could estimate the mass of the hidden Higgs field through both the VEV-scale and the Higgs' couplings.
A model of QFT was constructed in a renormalizable gauge, and we analyzed its perturbative structure.
After that, the radiative correction of the $X$-boson  propagator has been calculated at one loop approximation
to yield the Yukawa potential correction.
The form factors associated with the QED-vertex correction were calculated to confirm electron's anomalous
magnetic moment together with the computation of the interaction magnitude. The muon case was discussed.
Furthermore, we have introduced a renormalization group scheme to explore the current $X$-boson mass and its coupling constant
with the leptons of the Standard Model.

\end{abstract}


\maketitle

\pagestyle{myheadings}
\markright{The protophobic $X$-boson coupled to quantum electrodynamics}
%
%
%
%
%
%

\section{Introduction}

The anomalies of the excited state of $8$-Beryllium $(8\, \mbox{Be}^{\ast})$ relative to its ground state
have been revealed the existence of a new neutral $X$-boson through the nuclear decay $8 \, \mbox{Be}^{\ast} \rightarrow 8 \, \mbox{Be} + X$
\cite{KrasPRL2016}. Immediately, the $X$-boson decays into the electron-positron pair $X \rightarrow e^{+} \, + \, e^{-}$.
It has a vector feature like the photon one, but it must have a mass of approximately $m_{X}=17 \, \mbox{MeV}$.   
Moreover, it must mediate a weak force with a range of $12 \, fm$. In principle, its unification is associated with the introduction of an extra gauge symmetry $U'(1)$, besides the known gauge symmetry of the Standard Model (SM). 

The introduction of a new light particle via $U(1)$-extra is an old idea in physics which includes several scenarios of interactions beyond the SM \cite{WeinbergPRL78,Okun82,Holdom861,Holdom862,LangackerRMP2009}. Certainly,
the existence of a new boson can lead to the emergence of a fifth fundamental interaction in Nature \cite{GuHe2016}. In this context, the extended SM is based on the $SU_{c}(3)\times SU_{L}(2)\times U_{Y}(1) \times U'(1)$ gauge symmetry. For a complete review about the anomaly of beryllium decays, see
\cite{Feng2016}. Recently, a huge number of references show the alternative models to describe this extended SM \cite{Araki2017,Fayet2017,Ulrich2016,ChiangPLB2016,Seto2016,Kaneta2017}.
The effective Lagrangian that could describe the Abelian sector of the X-boson model can be written as
%
\begin{equation}\label{Leff}
{\cal L}_{eff}=-\frac{1}{4} F_{\mu\nu}^{\, 2}
- \frac{1}{4} X_{\mu\nu}^{\, 2}
+ \frac{\chi}{2} X_{\mu\nu} F^{\mu\nu}
+ \frac{1}{2} \, m_{X}^{\, 2} X_{\mu}^{\, 2} + J_{\mu} X^{\mu} \, ,
\end{equation}
where $X^{\mu\nu}$ is the field strength tensor of $X^{\mu}$, $F^{\mu\nu}$ is the corresponding one for the photon and
$J^{\mu}$ is the fermionic current coupled to the $X^{\mu}$-field
\begin{eqnarray}\label{Jcurrent}
J^{\mu}=\!\!\!\!\sum_{\Psi \, = \, e \, , \, u \, , \, d \, ,...} e \, \chi_{\Psi} \, \bar{\Psi} \, \gamma^{\mu} \, \Psi \; ,
\end{eqnarray}
where $\Psi$ represents any fermion of the SM, {\it i. e.}, $\Psi=\{ \, e \, , \, u \, , \, d \, , \, ... \, \}$.
The $X$-boson can also interact chirally with the SM leptons via axial current \cite{Feng2016,KozaczukPRD2017,KahnJHEP2017}.
%
%
The current in Eq. (\ref{Jcurrent}) defines the so called
protophobic interaction where the nucleons' coupling constant $(n,p)$ to the $X$-boson is multiplied by both the $\chi_{n}$
and $\chi_{p}$-parameters whose magnitude satisfies the inequality $\chi_{n} \gg \chi_{p}$.
This is also known as the millicharged interactions \cite{Holdom862}. We list some values of $\chi_{\Psi}$ for the electron,
neutron and neutrinos, following the $X$-boson phenomenology in the literature, namely,
\begin{eqnarray}\label{chiconstraints}
2 \times 10^{-4} < \!&|\chi_{e}|&\! < 1.4 \times 10^{-3} \; ,
\nonumber \\
|\chi_{n}| \!& < &\! 2.5 \times 10^{-2} \; ,
\nonumber \\
 \sqrt{|\chi_{\nu} \, \chi_{e}|} \!& \lesssim &\! 7 \times 10^{-5} \; .
\end{eqnarray}
By considering both the $u$- and $d$-quarks, the extreme protophobic limit $(\chi_{p}=0)$ parametrizes $\chi_{u}$ and $\chi_{d}$ since
\begin{eqnarray}
\chi_{u}\!\!&=&\!\!-\frac{\chi_{n}}{3} \simeq \pm \, 3.7 \times 10^{-3}
\nonumber \\
\chi_{d}\!\!&=&\!\!+\frac{2 \, \chi_{n}}{3} \simeq \mp \, 7.4 \times 10^{-3} \; .
\end{eqnarray}
%

In Eq. (\ref{Leff}), the $\chi$-parameter, computed in the range $10^{-6} < \chi < 10^{-3}$, mixes kinetically the boson $X^{\mu}$ with the usual electromagnetic (EM) photon $A^{\mu}$.  It is clear that the massive term spoils the $U'(1)$ symmetry.   The Lagrangian in Eq. (\ref{Leff}) has just the EM gauge symmetry $U(1)$. Therefore, a spontaneous symmetry breaking (SSB) mechanism spoils one of the Abelian symmetries to generate a mass value of $m_{X}=17 \, \mbox{MeV}$ in Eq. (\ref{Leff}). Consequently, the experimental value of $17 \, \mbox{MeV}$ defines the scale of a VEV, consequently we can estimate a mass range for the hidden Higgs.

In this paper, we will start with the known Abelian model of massless gauge fields based on a $U(1) \times U'(1)$ symmetry, with a kinetic mixing term
in the gauge sector. The hidden Higgs sector will be revisited in the $R_{\xi}$-gauge to spoil one of the $U(1)$ gauge symmetries, and therefore, we will perform the full diagonalization procedure to identify a massive gauge field from a VEV-scale as the $X$-boson. The remaining massless eigenstates will be identified as the EM-photon.   Hence, the model is a candidate to describe the interaction between leptons and the $X$-boson as well as the usual QED interaction with photon. This approach is similar to the $U(1)_{B}$ gauge boson together with kinetic mixing discussed in \cite{JFengPRD2017}.

Thus, we have a QED scenario coupled to the hypothetical fifth interaction associated with the $X$-boson mediator. Thereby, we can obtain a renormalizable and unitary model, where the interaction between both the $X$-boson and fermions of the SM satisfies the protophobic condition mentioned previously. We will discuss some aspects of the model from the point of view of the perturbative QFT with the motivation to confirm some constraints
on the $\chi$-parameters. For example, we can estimate the $X$-boson decay time by using the decay width of $X \rightarrow e^{+} \, + \, e^{-}$.
We obtain the differential cross section for electron-positron scattering $e^{+} \, e^{-} \, \rightarrow \, \mu^{+} \, \mu^{-}$
via $X$-boson at the tree level.
Moreover, we will analyze some aspects of the perturbation theory at one loop approximation: (i) The contribution of the $X$-boson to the
electron physical mass; (ii) The $X$-boson full propagator, the $m_{X}$-renormalized mass and the correspondent Uehling potential; (iii) The contribution of the $X$-boson into the QED-vertex calculation. Thus, the form factors yield a contribution of $\chi$-parameter to the electron's anomalous magnetic moment, that confirm the constraints in Eq. (\ref{chiconstraints}); (iv) Finally, we will obtain all the renormalization factors of the model by fixing the renormalization conditions.    In this way, the renormalization group scheme will be introduced to investigate both the current $X$-boson mass and its respective coupling constant as functions of an arbitrary dimensionless scale.
%
%

%
%
%
%
The organization of this paper obeys the following schedule: in section II we will discuss the model based on the $U(1) \times U'(1)$ symmetry
with a kinetic mixing.    The hidden Higgs sector will also be analyzed with a detailed full diagonalization to give a mass for the $X$-boson,
and the gauge sector in the $R_{\xi}$-gauge fixing.
In section III, we will calculate the decay rate of the $X$-boson into the electron-positron pair.
In section IV, we will obtain the scattering $e^{+} \, e^{-} \, \rightarrow \, \mu^{+} \, \mu^{-}$ at tree level.
In section V, the contribution of the $X$-boson to the electron self-energy will be calculated at one-loop.
In section VI, the correction to the $X$-boson full propagator will be used to obtain the correspondent Uehling Potential.
In section VII, the correction to the QED-vertex will be computed due to the interaction of the $X$-boson with the electron-positron
pair. In section VIII, we construct the model renormalization, and we will discuss the
functions that appear in the Callan-Symanzik equation. Finally, the conclusions and last remarks will be depicted in section IX.


\section{The Abelian model and the Higgs sector: New considerations}

In this section, we will restate the sector of gauge fields and leptons/quarks of the model governed by a
$U(1) \times U'(1)$ symmetry, with the kinect mixing $\chi$-parameter. This model is well known in the literature to describe
the hidden photon, or also, new gauge boson that could appear from the MeV-scale and below \cite{PospelovPRD2009}.
Here we will study in detail the diagonalization of the Higgs sector to identify the corresponding massive and massless eigenstates.

Let us begin with the gauge fields sector composed by two massless vector fields $\tilde{A}^{\mu}-\tilde{X}^{\mu}$ 
\begin{equation}\label{Lgauge}
{\cal L}_{gauge}= -\frac{1}{4} \, \tilde{\tilde{F}}_{\mu\nu}^{\; 2} -\frac{1}{4} \, \tilde{\tilde{X}}_{\mu\nu}^{\; 2}+\frac{\chi}{2} \, \tilde{\tilde{F}}_{\mu\nu} \, \tilde{\tilde{X}}^{\mu\nu} \; ,
\end{equation}
where $\chi$ is a real parameter that mixes the gauge fields of the model, and as usual, we define the field strength tensors
%
$\tilde{\tilde{F}}_{\mu\nu}=\partial_{\mu}\tilde{\tilde{A}}_{\nu}-\partial_{\nu}\tilde{\tilde{A}}_{\mu}
\hspace{0.3cm} \mbox{and} \hspace{0.3cm}
\tilde{\tilde{X}}_{\mu\nu}=\partial_{\mu}\tilde{\tilde{X}}_{\nu}-\partial_{\nu}\tilde{\tilde{X}}_{\mu}$.
%
%
%
%
The fermions sector is defined by the usual Lagrangian
\begin{equation}\label{Lleptons}
{\cal L}_{lept/quarks}=\bar{\Psi} \left( \, i \, \, \slash{\!\!\!\!D}-m \, {\uma} \, \right) \Psi \; ,
\end{equation}
where we have introduced the covariant derivative that couples leptons/quarks $\Psi=\left( \, e \, , \, \mu \, , \, \tau \, , \, u \, , \, d \, \right)$
to the gauge fields $\tilde{\tilde{A}}^{\mu}-\tilde{\tilde{X}}^{\mu}$ defined by
\begin{eqnarray}
D_{\mu} \Psi = \left( \phantom{\frac{1}{2}} \!\!\!\! \partial_{\mu}+ i \, Q \; g \, \tilde{\tilde{A}}_{\mu} + i \, Q' \, \varepsilon_{\Psi} \, g^{\prime} \, \tilde{\tilde{X}}_{\mu} \right) \Psi \, \,,
\end{eqnarray}
where $Q$ is the $U(1)$ charge generator, $Q'$ as the one corresponding to $U'(1)$, and $g$ and $g^{ \, \prime}$
are the coupling constants, respectively. The real parameter $\varepsilon_{\Psi}$ was introduced to characterize the interaction of the hidden
$X$ - boson with the leptons/quarks of the SM.
More explicitly, the interactions from (\ref{Lleptons}) between the fermions and the $\tilde{\tilde{A}}^{\mu}-\tilde{\tilde{X}}^{\mu}$ bosons are given by
\begin{eqnarray}\label{LintAY}
{\cal L}^{\, int} = - \, \bar{\Psi}\left( \, Q \; g \, \, \slash{\!\!\!\!\tilde{\tilde{A}}}+ Q' \, \varepsilon_{\Psi} \, g^{ \, \prime} \, \slash{\!\!\!\!\tilde{\tilde{X}}} \, \right) \Psi \; .
\end{eqnarray}
In fact, we did not identify the physical EM-photon and the X-boson in the model yet.
The physical fields are identified after a diagonalization procedure, with the help of the SSB mechanism.
%
%
%
%
%
%
%
Thereby, a hidden Higgs sector breaks one of the Abelian symmetries to give mass to the $X$-boson.
After this SSB, the EM-symmetry can be written such that
%
$U(1) \times U'(1) \stackrel{\langle \Phi \rangle}{\longmapsto} U_{em}(1)$.
%
To accomplish the task, the simplest scalar Lagrangian for a $\Phi$-Higgs is
\begin{eqnarray}\label{LHiggs}
{\cal L}_{Higgs}=
|D_{\mu} \Phi|^{2}-\mu^{\, 2} |\Phi|^{2} - \lambda \, |\Phi|^{4} \; ,
\end{eqnarray}
where $\mu$ and $\lambda$ are real parameters. The covariant derivative of the Lagrangian in Eq. (\ref{LHiggs}) acts on the $\Phi$-Higgs, coupling it to Abelian gauge fields
\begin{eqnarray}\label{DmuPhi1}
D_{\mu} \Phi= \left(\phantom{\frac{1}{2}} \!\!\!\! \partial_{\mu}+i \, Q_{\Phi} \, g \, \tilde{\tilde{A}}_{\mu} + i \, Q'_{\Phi} \, \varepsilon_{\Phi} \, g^{\prime}
\, \tilde{\tilde{X}}_{\mu} \, \right) \Phi \; .
\end{eqnarray}
The complex field $\Phi$ is a scalar singlet which transforms under the $U(1) \times U'(1)$ symmetry group.
%
%
The model reviewed is so represented by the Lagrangian $${\cal L}_{model}={\cal L}_{gauge}+{\cal L}_{lept/quarks}+{\cal L}_{Higgs}\,\,.$$
It is not difficult to verify that this model is invariant under the $U(1) \times U'(1)$ symmetry.
%
%

%
The minimal value of the Higgs potential can be obtained by using the non-trivial VEV of
the Higgs field which keeps the full invariance of the model. We choose it as the VEV constant
$\langle \Phi \rangle_{0}=v/\sqrt{2}$, where $v$ is the non-trivial VEV of the scalar field $\Phi$,
defined by $$v=\sqrt{-\frac{\mu^{2}}{\lambda}}\,\,,$$ when $\mu^{2}<0$.
We choose the standard parametrization of the $\Phi$-complex field as
\begin{eqnarray}\label{PhiGaugeparametrization}
\Phi(x)= \frac{v+H(x)}{\sqrt{2}} \, \, e^{ \, i \, \frac{\eta(x)}{v}} \; ,
\end{eqnarray}
where $H$ and $\eta$ are real functions, and $\eta$ is the Goldstone boson of the model. The VEV-$v$ defines a scale for the break
of the composite Abelian symmetry, where one of the gauge fields acquires a mass term. Therefore, after the SSB, the
Abelian sector is given by
\begin{eqnarray}\label{LGaugemassesXB}
{\cal L}_{gauge}\!\!\!\!\!&&=
-\frac{1}{4} \, \tilde{\tilde{F}}_{\mu\nu}^{\, 2}
-\frac{1}{4} \, \tilde{\tilde{X}}_{\mu\nu}^{\, 2}
+\frac{\chi}{2} \, \tilde{\tilde{F}}_{\mu\nu} \, \tilde{\tilde{X}}^{\mu\nu}
\nonumber \\
&&
\hspace{-0.8cm}
+\frac{v^2}{2} \left( \phantom{\frac{1}{2}} \!\!\!\!\! Q_{\Phi} \, g \, \tilde{\tilde{A}}_{\mu}+ Q'_{\Phi} \, \varepsilon_{\Phi} \, g^{\, \prime} \, \tilde{\tilde{X}}_{\mu} \right)^2
\nonumber \\
&&
\hspace{-0.8cm}
+\frac{1}{2}\left(\partial_{\mu}\eta\right)^{2}
+v \, \partial_{\mu}\eta \left(\phantom{\frac{1}{2}} \!\!\!\!\! Q_{\Phi} \, g \, \tilde{\tilde{A}}^{\mu} + \, Q'_{\Phi} \, \varepsilon_{\Phi} \, g^{\prime} \, \tilde{\tilde{X}}^{\mu} \right) .
\hspace{0.5cm}
\end{eqnarray}
%
%
The sector $\tilde{A}^{\mu}-\tilde{X}^{\mu}$ indicates a diagonalization procedure to obtain the mass of $X$-boson,
and the physical gauge bosons. To do it, we will write the $\tilde{A}^{\mu}-\tilde{X}^{\mu}$ Lagrangian in a matrix form
\begin{eqnarray}
{\cal L}_{\tilde{\tilde{A}}-\tilde{\tilde{X}}}=\frac{1}{2} \left(\tilde{\tilde{V}}^{\mu}\right)^{t} \! \Box \theta_{\mu\nu} K \, \tilde{\tilde{V}}^{\nu}
+ \frac{1}{2} \left(\tilde{\tilde{V}}^{\mu}\right)^{\!t}\! \eta_{\mu\nu} M^{2} \, \tilde{\tilde{V}}^{\nu} ,
\end{eqnarray}
where $(\tilde{\tilde{V}}^{\mu})^{t}=\left( \; \tilde{\tilde{A}}^{\mu} \; \; \tilde{\tilde{X}}^{\mu} \; \right)$, $K$
is the kinetic matrix
\begin{eqnarray}
K:=\left(
\begin{array}{cc}
1 & -\chi
\\
\\
- \chi & 1 \\
\end{array}
\right)
\; .
\end{eqnarray}
The mass matrix $M^{2}$ is given by
\begin{eqnarray}
M^{2}=v^2\left(
\begin{array}{cc}
Q_{\Phi}^2 \, g^2 & Q_{\Phi} \, Q'_{\Phi} \, g \, g' \, \varepsilon_{\Phi}
\\
\\
Q_{\Phi} \, Q'_{\Phi} \, g \, g' \, \varepsilon_{\Phi} & Q_{\Phi}'^{\,2} \, g'^{\,2} \, \varepsilon_{\Phi}^2
\end{array}
\right) \; .
\end{eqnarray}
We will carry out an $SO(2)$-orthogonal transformation
$\tilde{\tilde{V}} \; \longmapsto \; \tilde{\tilde{V}}= R \, \tilde{V} $, where $R^{t}\, R={\uma}$.
If we define the kinetic diagonal matrix as $K_{D}=R^{t} \, K \, R$, the eigenvalues of $K_{D}$ are given by
$1 \pm \chi$, the diagonal kinetic matrix is
\begin{eqnarray}
K_{D}=
\left(
\begin{array}{cc}
1 - \chi & 0
\\
\\
0 & 1 + \chi \\
\end{array}
\right) \; .
\end{eqnarray}
Thereby, the $\tilde{A}-\tilde{X}$ Lagrangian in terms of $\tilde{V}^{\mu}$ can be written as
\begin{equation}\label{LgaugeMatrizVtilKD}
{\cal L}_{\tilde{A}-\tilde{X}}= \frac{1}{2} \, \left(\tilde{V}^{\mu}\right)^{\, t} \Box \, \theta_{\mu\nu} \, K_{D} \, \tilde{V}^{\nu}
+ \frac{1}{2} \, \left(\tilde{V}^{\mu}\right)^{\, t} \eta_{\mu\nu} \, \tilde{M}^{2} \, \tilde{V}^{\nu} \; ,
\end{equation}
where $\tilde{M}^{2}=R^{t} \, M^{2} \, R$. Here, the solution for $R$ is the $SO(2)$-matrix with an angle of $45^{o}$
\begin{eqnarray}
R=\frac{1}{\sqrt{2}} \,
\left(
\begin{array}{cc}
1 & 1
\\
\\
-1 & 1 \\
\end{array}
\right) \; ,
\end{eqnarray}
so, the matrix mass $\tilde{M}^{2}$ is given by
\begin{equation}
\tilde{M}^{2}=\frac{v^2}{2}
\left(
\begin{array}{cc}
\left(Q_{\Phi} \, g-Q'_{\Phi} \, g' \varepsilon_{\Phi} \right)^2 & Q_{\Phi}^2 \, g^2-Q_{\Phi}'^2 \, g'^2 \varepsilon_{\Phi}^2
\\
\\
Q_{\Phi}^2 \, g^2-Q_{\Phi}'^2 \, g'^2 \varepsilon_{\Phi}^2 & \left(Q_{\Phi} \, g+Q'_{\Phi} \, g' \varepsilon_{\Phi}\right)^2 \\
\end{array}
\right) .
\end{equation}
We will write the matrix $K_{D}$ as $K_{D}=\left(K_{D}^{1/2}\right)^{t} \left( K_{D}^{1/2} \right)$
to include it into the kinetic term, redefining $\tilde{V} \, \longrightarrow \, K_{D}^{1/2} \, \tilde{V}$.
The solution for $K_{D}^{1/2}$ is
\begin{eqnarray}
K_{D}^{1/2}=
\left(
\begin{array}{cc}
\sqrt{1-\chi} & 0
\\
\\
0 & \sqrt{1+\chi}
\end{array}
\right) \; .
\end{eqnarray}
Thus, the Lagrangian in Eq. (\ref{LgaugeMatrizVtilKD}) can be rewritten as 
\begin{equation}\label{LgaugeMatrizVtilKD2}
{\cal L}_{\tilde{A}-\tilde{X}}= \frac{1}{2} \left(\tilde{V}^{\mu}\right)^{\, t} \Box \, \theta_{\mu\nu} \, \tilde{V}^{\nu}
+ \frac{1}{2} \left(\tilde{V}^{\mu}\right)^{\, t} \eta_{\mu\nu} \, M_{D}^{\, 2} \, \tilde{V}^{\nu} \, ,
\end{equation}
where the mass matrix is now $$M_{D}^{\, 2}=\left(K_{D}^{1/2}\right)^{-1} \tilde{M}^{2} \left(K_{D}^{1/2}\right)^{-1}\,\,,$$ namely,
\begin{eqnarray}\label{MDMass}
M_{D}^{2}=
\frac{v^2}{2}
\left(
\begin{array}{cc}
\frac{\left(Q_{\Phi} \, g-Q'_{\Phi} \, g' \varepsilon_{\Phi} \right)^2}{1-\chi} & \frac{Q_{\Phi}^2 \, g^2-Q_{\Phi}'^2 \, g'^2 \, \varepsilon_{\Phi}^2}{\sqrt{1-\chi^{2}}}
\\
\\
\frac{Q_{\Phi}^2 \, g^2-Q_{\Phi}'^2 \, g'^2 \, \varepsilon_{\Phi}^2}{\sqrt{1-\chi^{\, 2}}} & \frac{\left(Q_{\Phi} \, g+Q'_{\Phi} \, g' \varepsilon_{\Phi} \right)^2}{1+\chi} \\
\end{array}
\right) \; .
\end{eqnarray}
Since $M_{D}^{\, 2}$ is also symmetric, it can be diagonalized by an orthogonal matrix $S$, if we define $\tilde{V}^{\mu}=S \, V^{\mu}$,
whose rotation is
\begin{eqnarray}
\left(
\begin{array}{c}
\tilde{A}^{\mu}
\\
\\
\tilde{X}^{\mu} \\
\end{array}
\right)
=\left(
\begin{array}{cc}
\cos\theta & \sin\theta
\\
\\
-\sin\theta & \cos\theta \\
\end{array}
\right)
\left(
\begin{array}{c}
A^{\mu}
\\
\\
X^{\mu} \\
\end{array}
\right) \; .
\end{eqnarray}
We will end up with a fully diagonal Lagrangian such that
\begin{equation}\label{Lag3}
\mathcal{L}_{A-X} = \frac{1}{2} \left( V^{\mu} \right)^{t} \Box\theta_{\mu\nu} V^{\nu}
+ \frac{1}{2} \left(V^{\mu}\right)^{t} \eta_{\mu\nu} M_{diag}^2 V^{\nu} \; ,
\end{equation}
where $M_{diag}^{\, 2} = S^{t} \, M_{D}^{\, 2} \,  S$ is given by the eigenvalues of the mass matrix in Eq. (\ref{MDMass}), i. e.,
\begin{eqnarray}\label{MatrixMassaAX}
M_{diag}^{2}=
\left(
\begin{array}{cc}
0 & 0
\\
\\
0 & v^2 \, \frac{Q_{\Phi}^2 g^2 +Q_{\Phi}^2 g^{\prime 2} \varepsilon_{\Phi}^2 - 2 g \, g' Q_{\Phi} Q_{\Phi}' \, \chi \, \varepsilon_{\Phi}}{1-\chi^{2}} \\
\end{array}
\right) \, .
\end{eqnarray}
%
%
%
In this way, the $\theta$-mixing angle satisfies the relation
\begin{eqnarray}
\tan2\theta= \frac{\left(Q_{\Phi}^2 \, g^2-Q_{\Phi}^{'2}g^{\prime 2} \, \varepsilon_{\Phi}^2\right)\sqrt{1-\chi^{2}}}{2gg^{\prime}Q_{\Phi}Q_{\Phi}^{\prime} \, \varepsilon_{\Phi}- \chi\left( Q_{\Phi}^2 \, g^2+Q_{\Phi}^{'2}g^{\prime 2} \, \varepsilon_{\Phi}^2\right)}  \; .
\end{eqnarray}
The diagonal matrix in Eq. (\ref{MatrixMassaAX}) reveals the eigenstates of the gauge fields masses
$\left\{ \,  X^{\mu} , A^{\mu} \right\}$, where $A^{\mu}$ is massless, and $X^{\mu}$
acquires a squared mass that depends on the VEV scale. In fact, to identity $A^{\mu}$ as the EM-photon, we need to check
its interaction with the SM fermions.
%
%
%
Therefore, the transformation from the basis $\left\{ \, \tilde{\tilde{X}}^{\mu} \, , \, \tilde{\tilde{A}}^{\mu} \, \right\}$ to the basis of
physical $X$- and $A$-bosons $\left\{ \, X^{\mu} \, , \, A^{\mu} \, \right\}$ is represented by
$\tilde{\tilde{V}}=U \, V$, where $U=R \, (K_{D}^{1/2})^{-1} \, S$ is the matrix
\begin{eqnarray}
U=\frac{1}{\sqrt{2}}
\left(
\begin{array}{cc}
\frac{\cos\theta}{\sqrt{1+\chi}}-\frac{\sin\theta}{\sqrt{1-\chi}} & \frac{\cos\theta}{\sqrt{1-\chi}}+\frac{\sin\theta}{\sqrt{1+\chi}}
\\
\\
-\frac{\cos\theta}{\sqrt{1+\chi}}-\frac{\sin\theta}{\sqrt{1-\chi}} & \frac{\cos\theta}{\sqrt{1-\chi}}-\frac{\sin\theta}{\sqrt{1+\chi}} \\
\end{array}
\right) \; .
\end{eqnarray}
Notice that $U$ is not an $SO(2)$-matrix, since it has the determinant $\det{U}=\left(1-\chi^2\right)^{-1/2}$.
Thus, we obtain the transformation
\begin{eqnarray}\label{transfAX}
\tilde{\tilde{A}}^{\mu} \!&=&\! \left( \frac{\cos\theta}{\sqrt{1+\chi}}-\frac{\sin\theta}{\sqrt{1-\chi}} \right) \,
\frac{A^{\mu}}{\sqrt{2}}
\nonumber \\
&&
\hspace{-0.5cm}
+  \, \left( \frac{\cos\theta}{\sqrt{1-\chi}}+\frac{\sin\theta}{\sqrt{1+\chi}} \right)  \frac{X^{\mu}}{\sqrt{2}}
\nonumber \\
\tilde{\tilde{X}}^{\mu} \!&=&\! -\left( \frac{\cos\theta}{\sqrt{1+\chi}}+\frac{\sin\theta}{\sqrt{1-\chi}} \right)  \, \frac{A^{\mu}}{\sqrt{2}}
\nonumber \\
&&
\hspace{-0.5cm}
+\left( \frac{\cos\theta}{\sqrt{1-\chi}}-\frac{\sin\theta}{\sqrt{1+\chi}} \right)  \, \frac{X^{\mu}}{\sqrt{2}} \; .
\end{eqnarray}
%
%
The simplest solution for the diagonalization of the mass matrix in Eq. (\ref{MDMass}) is to choose that $Q_{\Phi}=0$ and
so, the $\theta$-mixing angle satisfies the relation $$\tan\,(2\theta)=\sqrt{1-\chi^2}/\chi\,\,,$$ and the transformations in Eq. 
(\ref{transfAX}) reduces to 
\begin{eqnarray}\label{shiftAX}
\tilde{\tilde{A}}^{\mu} \!&=&\! A^{\mu} + \frac{ \chi \, X^{\mu}}{\sqrt{1-\chi^2}}
\nonumber \\
\tilde{\tilde{X}}^{\mu} \!&=&\! \frac{X^{\mu}}{\sqrt{1-\chi^2}} \; .
\end{eqnarray}
Hence, the non-trivial eigenvalue of the mass matrix in Eq. (\ref{MatrixMassaAX}) is connected to the mass of $X$-boson
\begin{eqnarray}\label{massC}
m_{X}= \frac{|Q_{\Phi}' \, \varepsilon_{\Phi}| \, g' \, v}{\sqrt{1-\chi^2}} \; .
\end{eqnarray}
Hence,
we can parametrize the coupling constants such as $g=g^{\prime}=e$, and consequently, $Q_{em}^{\, (\Phi)}=Q_{\Phi}=0$.
The hidden Higgs charge $Q'$ is standardly chosen in the literature as being equal to $Q'_{\Phi}=+3$.   Thus, the $X$-mass is
$m_{X}=3 \,|\varepsilon_{\Phi}| \, e \, v/\sqrt{1-\chi^2}$.
In particular, this value for the hidden charge is associated with the instability of exotic fermions that can be introduced in the model,
and it would be bounded based on a dark matter scenario, for more details, see \cite{KitaharaPRD2017,Duerr2014,Duerr2015}.

Substituting the shift in Eq. (\ref{shiftAX}) into Eq. (\ref{LintAY}), the interaction in the basis of $\left\{ \, X^{\mu} \, , \, A^{\mu} \, \right\}$
is 
%
%
\begin{eqnarray}\label{LintCY}
{\cal L}^{\, int} = - \, e \, Q_{em} \, \bar{\Psi} \, \, \slash{\!\!\!\! A} \, \, \Psi
- \, e \, Q_{X} \, \bar{\Psi} \, \, \slash{\!\!\!\!X} \, \, \Psi \; ,
\end{eqnarray}
where the electric charge is $Q_{em}=Q$, and the protophobic charge generator $Q_{X}$ is defined by
\begin{eqnarray}\label{chil}
Q_{X}\,=\, \, Q_{em} \, \tilde{\chi}+ Q' \, \tilde{\varepsilon}_{\Psi} \; ,
\end{eqnarray}
where $\tilde{\chi}$ and $\tilde{\varepsilon}_{\Psi}$ are, respectively, given by
\begin{eqnarray}
\tilde{\chi}=\frac{\chi}{\sqrt{1-\chi^{2}}}
\hspace{0.3cm} \mbox{and} \hspace{0.3cm}
\tilde{\varepsilon}_{\Psi}= \frac{\varepsilon_{\Psi}}{\sqrt{1-\chi^{2}}} \; .
\end{eqnarray}
%
%
%
%
%
%
Thus, the previous gauge invariant Lagrangian is reduced to
\begin{eqnarray}\label{LGaugemassesAC}
{\cal L}_{gauge}& =&
-\frac{1}{4} F_{\mu\nu}^{\, 2}
-\frac{1}{4} X_{\mu\nu}^{\, 2}
+\frac{1}{2} m_{X}^{\, 2} X_{\mu}^{\, 2}
+\frac{1}{2} \left(\partial_{\mu}\eta\right)^{2} \nonumber \\
&+&\frac{1}{2}  m_{X} \partial_{\mu}\eta \, X^{\mu} \, .
\end{eqnarray}
%
%
%
To eliminate the mixed last term $\eta-X^{\mu}$, we can add a gauge fixing Lagrangian
\begin{eqnarray}\label{LgfAChi}
{\cal L}_{gf}= -\frac{1}{2\alpha} \left( \partial_{\mu}A^{\mu}\right)^{2}
-\frac{1}{2\beta}\left( \partial_{\mu}X^{\mu} - \frac{\beta}{2} \, m_{X} \, \eta \right)^{2} ,
\end{eqnarray}
where $\{\alpha,\beta\}$ are real parameters. The surface terms can be eliminated
if we imagine the Lagrangian integrated throughout the entire space-time, so we can find all the gauge sector terms
\begin{eqnarray}\label{FreeLACZW}
{\cal L}_{gauge+gf}\!\!&=&\!\!-\frac{1}{4} \, F_{\mu\nu}^{\, 2}
-\frac{1}{2\alpha} \left(\partial_{\mu}A^{\mu} \right)^2
\nonumber \\
&&
\hspace{-0.5cm}
-\frac{1}{4} \, X_{\mu\nu}^{\, 2}-\frac{1}{2\beta} \left(\partial_{\mu}X^{\mu}\right)^{2}
+\frac{1}{2} \, m_{X}^{\, 2} X_{\mu}^{\, 2} \; ,
\end{eqnarray}
%
where, again, $F^{\mu\nu}$ is the EM-Maxwell strength field tensor and
$X^{\mu\nu}=\partial^{\mu}X^{\nu}-\partial^{\nu}X^{\mu}$
is the correspondent $X$-boson one.
%
%
%
%
%
The Higgs sector of the VEV is represented by the Lagrangian
\begin{eqnarray}
{\cal L}_{Higgs} \!&=&\! \frac{1}{2} \, \left( \partial_{\mu} H \right)^{2} - \frac{1}{2} \, m_{H}^{\, 2} \, H^{2}
-\frac{\lambda}{4} \, H^{4}-\lambda \, v \, H^{3}
\nonumber \\
&&
\hspace{-0.5cm}
+\,9 \, \varepsilon_{\Phi}^2 \, e^2 \, v \, H \, X_{\mu}^{\,2}+\frac{9}{2} \, \varepsilon_{\Phi}^2 \, e^2 \, H^2 \, X_{\mu}^{\,2}
\, ,
\end{eqnarray}
where the Higgs mass is identified by $m_{H}=\sqrt{2 \, \lambda \, v^2}$. All the propagators are well defined at the ultraviolet range, {\it i. e.},
they are renormalizable and the model is unitary too. The inversion of the quadratic Lagrangian in Eq. (\ref{FreeLACZW}) gives the renormalized
$X$-propagator in the momentum space
\begin{eqnarray}\label{PropAA}
\langle X_{\mu} X_{\nu} \rangle \!\!&=&\!\! -\frac{i}{k^{2}-m_{X}^{\, 2}}\left[ \phantom{\frac{1}{2}} \!\!\! \eta_{\mu\nu}
+\left(\beta-1 \right) \frac{k_{\mu} \, k_{\nu}}{k^{2}-\beta\, m_{X}^{\, 2}} \right] .
\hspace{0.5cm}
\end{eqnarray}
%
%
%

Now, we will return to the analysis of the hidden $X$-charge in Eq. (\ref{chil}).   This is exactly the model with the charge $Q_{X}$
introduced in \cite{JFengPRD2017}, but with the $R_{\xi}$-gauge fixing useful during the loop calculation.   Firstly, we will analyze the leptons' case where 
the $Q_{X}$-charge has the generators $Q_{em}=-1$ and $Q'=0$, so we obtain the relation
\begin{eqnarray}
Q_{X}^{(e)}=Q_{X}^{(\mu)}=Q_{X}^{(\tau)}=- \, \tilde{\chi} \; .
\end{eqnarray}
Therefore, the $X$-interaction with leptons has a universal coupling constant, namely, $- \, e \, \chi$.
We will calculate the $\chi$-value using the anomalous magnetic moment through the vertex correction at one loop.
%
%
%
%
%
To compute the $\tilde{\varepsilon}$-parameter, we have to analyze the charges in the EM interaction  scenario with $up$- and $down$-quarks.
In this case, the $\tilde{\varepsilon}_{\Psi}$-parameter in Eq. (\ref{chil}) for up and down-quarks, respectively, are given by
\begin{eqnarray}
Q_{X}^{(u)} \!&=&\! +\frac{2}{3} \, \tilde{\chi}+\frac{1}{3} \, \tilde{\varepsilon}_{u}
\qquad 
\mbox{and} \nonumber \\
Q_{X}^{(d)} \!&=&\! -\frac{1}{3} \, \tilde{\chi}+\frac{1}{3} \, \tilde{\varepsilon}_{d} \; \;,
\end{eqnarray}
where $Q'=+1/3$ has been chosen conveniently. Using the relation for nucleon charges $\chi_{n}=\chi_{u}+2 \, \chi_{d}$ together with $\tilde{\varepsilon}_{u}=\tilde{\varepsilon}_{d}=\tilde{\varepsilon}$ for the first generation, we obtain that
$\chi_{n}=\tilde{\varepsilon} < 2.5 \times 10^{-2}$. Thus, the $\tilde{\chi}$- and $\tilde{\varepsilon}$-parameters which emerge
from the model are constrained by experimental results.
The entire model that includes quarks is the one that unites the former to the weak interaction through an extended $SU_{L}(2) \times U_{Y}(1) \times U'(1)$ symmetry.
Furthermore, the currents can have vector and axial components into the interactions between the $X$-boson and the SM fermions.
The interactions between quarks in the radiative corrections will not be considered in this paper. As usual, we will represent the wavy lines as representing the EM photon, while the $X$-boson will be described by the coil lines at the diagrams pictured just below. The vertex for the interaction between the $X$-boson and leptons of the SM in Eq. (\ref{LintCY}) is given by 
\begin{figure}[!h]
\begin{center}
\newpsobject{showgrid}{psgrid}{subgriddiv=1,griddots=10,gridlabels=6pt}
\begin{pspicture}(5,1)(7.5,2.5)
\psset{arrowsize=0.2 2}
\psset{unit=0.8}
%
%
\pscoil[coilarm=0,coilwidth=0.2,coilheight=1.0,linecolor=black](6.5,1.05)(6.5,3)
\psline[linecolor=black,linewidth=0.5mm]{-}(5,1)(8,1)
\psline[linecolor=black,linewidth=0.5mm]{->}(5,1)(6,1)
\psline[linecolor=black,linewidth=0.5mm]{->}(7,1)(7.55,1)
\put(6.8,2.8){\large$X^{\mu}$}
\put(4.9,1.2){\large$\bar{\Psi}$}
\put(7.7,1.2){\large$\Psi$}
\put(8.7,1){\large$\Gamma_{X}^{\, \mu}=- \, i \, e \, \chi \, \gamma^{\mu}.$}
\end{pspicture}
%
%
%
\end{center}
\end{figure}
%
%
%

\noindent
Using the value $m_{X}=17 \, \mbox{MeV}$, the $X^{\mu}$-boson mass fixes the VEV-scale at
\begin{eqnarray}\label{massX}
v \simeq \frac{19}{|\tilde{\varepsilon}_{\Phi}|} \, \, \mbox{MeV} \simeq \, 6.3 \, \, \mbox{GeV} \; .
\end{eqnarray}
%
%
%
In this way,
the dark Higgs mass can be calculated within the range of the GeV-scale. It must be lighter than
the usual Higgs which has mass equal to $125 \, \mbox{GeV}$.
%
%
%
%
%
%
%
%
Therefore, we have gotten a similar QFT model consistent with both the requirements
of renormalization and unitarity, and also with the $X$-boson phenomenology.
Besides, the interaction sector of $X$-boson with the SM's leptons/quarks satisfies
the experimental constraints with $\varepsilon$-parameter. In the next section,
we will analyze the decay width to compute the time decay of the $X$-boson.



\section{The Decay time of the X-boson}

The $X$-boson phenomenology starts with the description of two decay processes. One of them
is the $\pi^{0}$-decay into a massive dark photon with the coupling to SM particles that
are proportional to their electric charge \cite{GuHe2016}.    In an X-boson context,
the neutral pion decay is $\pi^{0} \rightarrow X \, \, \gamma$ with the bound for
$\chi_{p}$-proton parameter around the $|\chi_{p}| \lesssim 0.8-1.2 \times 10^{-3}$.
An $SU(2)$-model that includes the scalar pions can be the best description of this process.
The second decay can be found in the Atomki pair spectrometer experiment that observes the excited
$8$-beryllium decay $8Be^{\star} \, \rightarrow \, 8Be + X $ followed by $X \, \rightarrow \, e^{+} \, e^{-}$
\cite{KrasPRL2016,Gul2016}. Thus, the model proposed here has the framework of interaction of the X-boson with $e^{\pm}$-pair proportional
to the fundamental charge. Following the diagrammatic notation, the $X$-boson decaying into the $e^{+} \, \, e^{-}$ pair is represented by the vertex diagram in Figure 1,
%
%
\begin{figure}[!h]
\begin{center}
\newpsobject{showgrid}{psgrid}{subgriddiv=1,griddots=10,gridlabels=6pt}
\begin{pspicture}(0,-0.3)(7,3.43)
\psset{arrowsize=0.2 2}
\psset{unit=0.8}
%
%
%
\pscoil[coilarm=0,coilwidth=0.5,coilheight=1.0,linecolor=black](1,2)(5,2)
\psline[linecolor=black,linewidth=0.7mm]{->}(5,2)(6.5,3.5)
\psline[linecolor=black,linewidth=0.7mm]{-}(5.8,2.8)(7.5,4.5)
\psline[linecolor=black,linewidth=0.7mm](6,3)(5,2)
\psline[linecolor=black,linewidth=0.7mm](5,2)(6.2,0.8)
\psline[linecolor=black,linewidth=0.7mm]{<-}(6,1)(7.5,-0.5)
%
%
%
\put(3,2.6){\Large$X^{\mu}$}
%
\put(3,1){\Large$k^{\mu}$}
\put(6,3.6){\large$e^{-}$}
\put(6,-0.1){\large$e^{+}$}
\put(6.6,3.1){$(p,s)$}
\put(6.6,0.7){$(p^{\prime},s^{\prime})$}
%
%
%
\end{pspicture}
\caption{The vertex diagram for the $X$-boson decay into an $e^{+} \, e^{-}$ pair.}\label{figvm}
\end{center}
\end{figure}
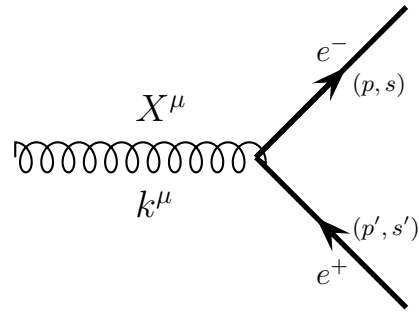

\noindent

Using the usual rules of QFT, the decay rate is given by the expression
\begin{equation}
\Gamma=\frac{1}{2\pi^{2}}\frac{1}{2 k^{0}} \int\frac{d^{3}{\bf p}}{2p^{0}} \int\frac{d^{3}{\bf p}'}{2 p^{\prime 0}} \,
\delta^{4}(k-p-p') \, \frac{1}{12}\sum_{\lambda,s,s'}|\mathcal{M}|^{2} \; ,
\end{equation}
where $k^{\mu}$ is the X-boson four-momentum, and $p^{\mu}$ and $p^{\prime\mu}$ are the external momenta of the electron and positron, respectively.
The electron-positron elastic scattering amplitude is
\begin{equation}
i\mathcal{M}\left( X \, \rightarrow \, e^{+} \, e^{-} \right)=\epsilon_{\mu}(k,\lambda) \, \bar{u}(p,s) \, \left(-ie\chi\gamma^{\mu}\right) \, v(p',s') \; ,
\end{equation}
where $\epsilon^{\mu}$ is the polarization vector, and $u$ and $v$ represent the wave plane amplitudes.
Using the completeness relation
\begin{eqnarray}
\sum_{\lambda}\epsilon_{\mu}(k,\lambda) \, \epsilon_{\nu}(k,\lambda)\!&=&\!-\eta_{\mu\nu}+\frac{k_{\mu} \, k_{\nu}}{m_{X}^{\,2}} \; ,
\nonumber \\
\sum_{s}u_{\alpha}(p,s) \, \bar{u}_{\beta}(p,s)\!&=&\!\left(\slash{\!\!\!p}+m\right)_{\alpha\beta} \; ,
\nonumber \\
\sum_{s'}v_{\alpha}(p',s') \, \bar{v}_{\beta}(p',s')\!&=&\!\left(\slash{\!\!\!p}\,'-m\right)_{\alpha\beta} \; ,
\end{eqnarray}
the $\Gamma$-decay factor assumes the form
\begin{eqnarray}
&&\Gamma=\frac{e^{2} \chi^{\, 2}}{12 \pi^{2} \, m_{X}^{3}} \! \int \! \frac{d^{3}{\bf p}}{2p^{\, 0}} \, \frac{d^{3}{\bf p}'}{2p'^{\, 0}} \, \delta^{4}\left(k-p-p'\right)
\nonumber \\
&&
\times \left[\left(p \cdot{k} \right) \left(p' \cdot{k}\right)
+\frac{1}{4} \, m_{X}^{\, 2} \left( m_{X}^{\, 2}+4 \, m^{2} \right) \right] ,
\hspace{0.8cm}
\end{eqnarray}
where we have used the on-shell condition $k^{2}=m_{X}^{\, 2}$. Solving the above integral, we arrive at the decay rate
\begin{equation}
\Gamma\left( \, X \rightarrow e^{+} \; e^{-} \, \right)=\frac{\alpha \, \chi^{\, 2}}{3}\, m_{X}
\sqrt{1-4\, \frac{m^{2}}{m_{X}^{\, 2}}} \left( \, 1
+2 \, \frac{m^{2}}{m_{X}^{\, 2}} \, \right)
\; ,
\end{equation}
where $\alpha=e^2/4\pi\simeq1/137$ is the fine structure constant, and $m_{X}> 2 \, m$.
%
%
%
%
The decay $X \rightarrow e^{+} \, \, e^{-}$ can be approximated by $m_{X} \gg 2 \, m$,
so the $\Gamma$-factor is given by
\begin{equation}
\Gamma\left( \, X \rightarrow e^{+} \; e^{-} \, \right) \approx \frac{\alpha \, \chi^{\, 2}}{3} \, m_{X} \; .
\end{equation}
Using $m_{X}= 17 \, \mbox{MeV}$, and the constraints from Eq. (\ref{chiconstraints}), we obtain the range
\begin{equation}
1.6 \times 10^{-9} \, \mbox{MeV} < \Gamma\left( \, X \rightarrow e^{+} \; e^{-} \, \right) < 8 \times 10^{-8} \, \mbox{MeV} \, .
\end{equation}
Other possible process is the $X$-decay into light neutrinos, i. e., $X \rightarrow \bar{\nu} \, \nu$. Following this framework, the decay width is
\begin{equation}
\Gamma\left( \, X \rightarrow \bar{\nu} \; \nu \, \right) \approx \frac{\alpha \, \chi_{\nu}^{\, 2}}{3} \, m_{X} \; .
\end{equation}
Thus, for the constraint in Eq. (\ref{chiconstraints}), we can obtain the range
\begin{equation}
5 \times 10^{-13} \, \mbox{MeV} < \Gamma\left( \, X \rightarrow \bar{\nu} \; \nu \, \right) < 2.3 \times 10^{-10} \, \mbox{MeV} \, .
\end{equation}
Hence, the mode lifetime $\tau$ is given by
\begin{equation}
\tau=\frac{1}{\Gamma\left( \, X \rightarrow e^{+} \; e^{-} \, \right)+\Gamma\left( \, X \rightarrow \bar{\nu} \; \nu \, \right)} \; ,
\end{equation}
namely, it has the range\footnote{Here, we have used the conversion formula $1 \, \mbox{MeV}=1.52 \times 10^{21} \, \mbox{s}^{-1}$ in the natural units $\hbar=c=1$. }.
\begin{eqnarray}
8.3 \times 10^{-15} \, \mbox{s} < \tau < 4.2 \times 10^{-12} \, \mbox{s} \; .
\end{eqnarray}
%
%
%
%
However, the better way to investigate this process is through a model $SU_{L}(2)\times U_{Y}(1) \times U'(1)$ which includes the
left-handed neutrinos (and the right-components), where its interaction with the $X$-boson must depend on mixing angles, like the
$\theta_{W}$-Weinberg angle, for example. This enlarged framework with the weak interaction sector is an ongoing research.
%

%

\section{The scattering $e^{+} \, e^{-}  \stackrel{X}{\rightarrow} \mu^{+} \, \mu^{-}$}

The measurements of a neutron-nucleus scattering motivate us to explore the Yukawa potential acting on the $X$-boson,
and the corresponding scattering process \cite{Barbieri75}. In the context of this model, we can use the $e^2$-leading
order for the electron-positron scattering process $e^{+} \, e^{-}  \stackrel{X}{\rightarrow} \mu^{+} \, \mu^{-}$ to obtain
its scattering cross-section. The process is so represented by the diagram in Figure 2 below
%
%
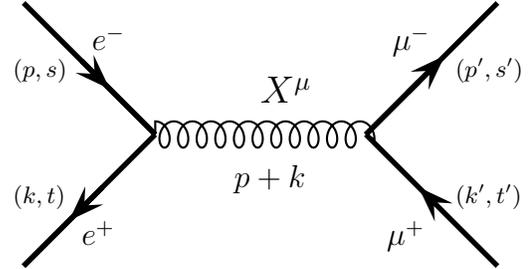
\begin{figure}[!h]
\begin{center}
\newpsobject{showgrid}{psgrid}{subgriddiv=1,griddots=10,gridlabels=6pt}
\begin{pspicture}(-2,-0.3)(8,2.95)
\psset{arrowsize=0.2 2}
\psset{unit=0.7}
%
%
\pscoil[coilarm=0,coilwidth=0.5,coilheight=1.0,linecolor=black](1,2)(5,2)
\psline[linecolor=black,linewidth=0.7mm]{->}(5,2)(6.5,3.5)
%
%
\psline[linecolor=black,linewidth=0.7mm]{-}(5.8,2.8)(7.5,4.5)
\psline[linecolor=black,linewidth=0.7mm](6,3)(5,2)
\psline[linecolor=black,linewidth=0.7mm](5,2)(6.2,0.8)
\psline[linecolor=black,linewidth=0.7mm]{<-}(6,1)(7.5,-0.5)
%
%
\psline[linecolor=black,linewidth=0.7mm,ArrowInside=->,ArrowInsidePos=0.6](-1.5,4.5)(1,2)
\psline[linecolor=black,linewidth=0.7mm,ArrowInside=->,ArrowInsidePos=0.6](1,2)(-1.5,-0.5)
\psline[linecolor=black,linewidth=0.7mm]{-}(0,1)(-1.5,-0.5)
\psline[linecolor=black,linewidth=0.7mm]{-}(1,2)(-0.4,3.4)
%
%
%
\put(3,2.6){\Large$X^{\mu}$}
%
\put(2.5,1){\large$p+k$}
\put(5.5,3.6){\large$\mu^{-}$}
\put(5.4,-0.1){\large$\mu^{+}$}
\put(6.7,3.1){$(p',s')$}
\put(6.7,0.7){$(k',t')$}
\put(-0.2,3.6){\large$e^{-}$}
\put(-0.4,-0.1){\large$e^{+}$}
\put(-1.7,3.1){$(p,s)$}
\put(-1.7,0.7){$(k,t)$}
%
%
%
%
\end{pspicture}
%
%
\caption{The scattering diagram $e^{+} \, e^{-}  \stackrel{X}{\rightarrow} \mu^{+} \, \mu^{-}$
involving the interaction of $X$-boson with the electron-positron pair at the tree level.}
\label{ScatteringX}
\end{center}
\end{figure}

\begin{figure}[h]
\centering
\includegraphics[scale=0.43]{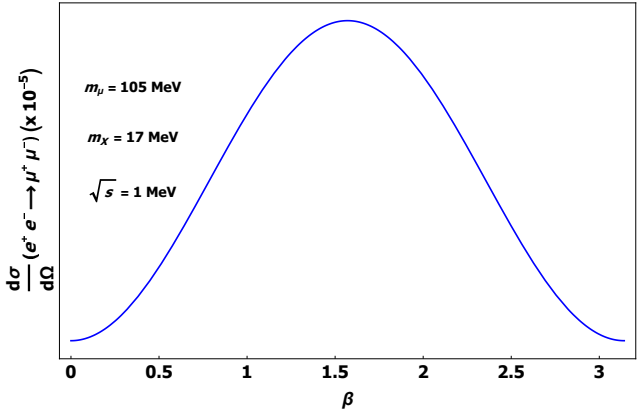}
\caption{The differential cross section of $e^{+} \, e^{-} \rightarrow \mu^{+} \, \mu^{-}$
as function of the $\beta$-angle. We adopt the values of
$m_{X}=17 \, \mbox{MeV}$, $m_{\mu}=105 \, \mbox{MeV}$
and $\sqrt{s}=1 \, \mbox{MeV}$ for the CM frame energy.} \label{Figdsigma}
\end{figure}

Using the rules of the diagram above, the amplitude for this scattering at the 
tree level is\footnote{We use the $X$-boson propagator in the Feynman gauge $(\beta=1)$.}
\begin{eqnarray}
&&
i{\cal M}\left(e^{+} \, e^{-}  \stackrel{X}{\rightarrow} \mu^{+} \, \mu^{-} \right)
=\bar{v}\left(k,t\right)
\left(-ie\chi\gamma^{\mu}\right)
u\left(p,s\right) 
\nonumber \\
&&
\times \,
\frac{-i\eta_{\mu\nu}}{(p+k)^{2}-m_{X}^{2}} \,
\bar{u}\left(p^{\prime},s^{\prime}\right)
\left(-ie\chi\gamma^{\nu}\right) \, v\left(k^{\prime},t^{\prime}\right) .
\end{eqnarray}
where the momentum are $p$ and $k$ for electron-positron pair, and $p'$ and $k'$ for the muon pair $\mu^{-}$ and $\mu^{+}$.
Let us consider the collision in the center-of-mass (CM) frame of the $e^{+} \, e^{-}$ pair. In this case,
${\bf p}+{\bf k}={\bf p}'+{\bf k}'={\bf 0}$ and we denote the electron and muon energy as being
$s=(p+k)^2=(p'+k')^2$. Using the rules of QFT, the differential cross section is given by
\begin{eqnarray}\label{dsigma}
\frac{d\sigma}{d\Omega} \hspace{-0.4cm}&&\left( e^{+} \, e^{-}  \stackrel{X}{\rightarrow} \mu^{+} \, \mu^{-} \right)=\frac{\alpha^{2}\chi^4}{4\left(s-m_{X}^{2}\right)^{2}} 
\nonumber \\
&&
\hspace{-0.3cm}
\times \,
\left[ \phantom{\frac{1}{2}} \hspace{-0.25cm} s\left(1+\cos^2\beta\right)
+8 \, m_{\mu}^{2} \, \sin^2\beta \right] \, ,
\end{eqnarray}
where $\beta$ is the scattering angle between the tri-momentum ${\bf p}$ and ${\bf p}'$,
or between ${\bf k}$ and ${\bf k}'$ in the CM frame. We also consider $m_{e}\approx 0$
when compared to the muon mass $m_{\mu}=105.7 \, \mbox{MeV}$.
The result in Eq. (\ref{dsigma}) is illustrated in Figure \ref{Figdsigma}
as a function of the $\beta$-angle.
Using the value $\sqrt{s}=1 \, \mbox{MeV}$ for the CM energy and the previous masses,
the total cross section of this process is
\begin{eqnarray}
\sigma\hspace{-0.4cm}&&\left( e^{+} \, e^{-}  \stackrel{X}{\rightarrow} \mu^{+} \, \mu^{-} \right)=
\pi\alpha^{2}\chi^4 \,
\frac{s+8m_{\mu}^{2}}{\left(s-4m_{X}^{2}\right)^{2}}
\nonumber \\
&&
=1.78 \times 10^{-4} \, \chi^{4} \, \mbox{MeV}^{-2} \; .
\end{eqnarray}
\section{The electron self-energy with the X-boson correction}

Over the last sections, we have obtained some phenomenological results for the Abelian model of the $X$-boson
at the leading order in a perturbative QFT. From now on, we will discuss the radiative corrections
at one loop approximation. The first one is due to the electron propagator, where we have investigated the influence of the
X-boson mass in the physical electron mass. Using the previous rules, the electron propagator at one loop can be
%
written as the sum of usual QED contribution and the self-energy due to the $X$-boson propagator
%
%
%
%
\begin{eqnarray}
\Sigma_{1}\left(\slash\!\!\!p\right) = \Sigma_{1}^{QED}\left(\slash\!\!\!p \right)
\!+\!\Sigma_{1}^{\,(X)}\left(\slash\!\!\!p \right) \; .
\end{eqnarray}
The expressions of $\Sigma_{1}^{QED}\left(\slash\!\!\!p \right)$ and $\Sigma_{1}^{\,(X)}\left(\slash\!\!\!p \right)$
are represented by the integrals
\begin{equation}\label{Sigma1QED}
\Sigma_{1}^{QED}\!\left(\slash\!\!\!p \right)=- i \, e^2
\int \frac{d^{4}k}{(2\pi)^{4}} \,
\frac{\gamma^{\mu}\left(\slash\!\!\!k+\slash\!\!\!p+m\right)\gamma_{\mu}}{\left[ \left( k+p \right)^{2}-m^{2} \right]\left(k^{2}-m_{\gamma}^{2}\right)} \; ,
\end{equation}
and
\begin{equation}
\Sigma_{1}^{\,(X)}\!\left(\slash\!\!\!p \right)=- i \, e^2 \, \chi^2
\int \frac{d^{4}k}{(2\pi)^{4}} \,
\frac{\gamma^{\mu}\left(\slash\!\!\!k+\slash\!\!\!p+m\right)\gamma_{\mu}}{\left[ \left( k+p \right)^{2}-m^{2} \right]\!\left( k^{2}-m_{X}^{\,2} \right)} \, .
\end{equation}
where $m_{\gamma}$ is the photon mass introduced as a regulator
to control the infrared divergence that emerges from Eq. (\ref{Sigma1QED}) when $m_{\gamma}=0$. It is clear that after the renormalization,
we must withdraw the $m_{\gamma}$-parameter since we have a possible limit such as $m_{\gamma} \rightarrow 0$.
Therefore, we have two similar integrals where the $\chi$-factor is less than one.
These integrals have linear divergences in the ultraviolet region,
so we will use the dimensional regularization to control the divergences, {\it i. e.},
the integral dimension is altered by a $\omega$-regularization parameter $4 \rightarrow 2\omega$, where the physical dimension is 
recovered, obviously, when $\omega \rightarrow 2$. In this way, the coupling constant is redefined to keep it dimensionless, i. e.,
$e \, \rightarrow \, e \, \mu^{2-\omega}$, where $\mu$ is an arbitrary energy scale. In fact, when we withdraw the regularization parameter,
the divergent term will be isolated from the physical term, which makes the expansion $\omega=2-\varepsilon$, for $\varepsilon \rightarrow 0$.
Therefore, the QED-contribution in the Feynman gauge is known in the literature as the result below
\begin{eqnarray}
&&
\, \Sigma_{1}^{QED}(\slash\!\!\!p \, , \, \varepsilon)= \frac{\alpha}{4\pi}
\left(-\slash\!\!\!p+4 m \right) \frac{\mu^{2\varepsilon}}{\varepsilon}
-\frac{\gamma \, \alpha}{4\pi} \left( - \slash\!\!\!p \, + 4 m \right)
\nonumber \\
&&
- \frac{\alpha}{2\pi} \, \int_{0}^{1} \! dz \left[ \left( 1-z \right) \, \slash\!\!\!p \, - 2 \, m \right]
\ln\left[ \frac{4\pi \mu^{2}}{m^{2} \, z-p^{2} \, z \, (1-z)} \right] \; ,
\nonumber \\
\end{eqnarray}
where $p^{2}<2\, m^2$, and $\gamma\simeq 0.57$ is the Euler-Mascheroni constant. The regularized contribution of the $X$-boson is given by the integral
\begin{eqnarray}\label{SigmaXint}
&&-i \, \Sigma_{1}^{\, (X)}(\slash\!\!\!p \, , \, \omega ) =
- e^2 \, \chi^2 \, ( \mu^{2} )^{2-\omega} \times
\nonumber \\
&&
\times \, \int \frac{d^{2\omega}k}{(2\pi)^{2\omega}} \,
\frac{\gamma^{\mu}\left(\slash\!\!\!k+\slash\!\!\!p+m\right)\gamma_{\mu}}{\left[ \left( k+p \right)^{2}-m^{2} \right]\left(  k^{2}-m_{X}^{\,2} \right)} \; .
\hspace{0.8cm}
\end{eqnarray}
Using the technique given in the literature concerning the Feynman integrals, the result of the one in Eq. (\ref{SigmaXint}), for $\omega=2-\varepsilon$ is
%
%
%
%
%
%
%
%
\begin{eqnarray}\label{SigmaX}
\Sigma_{1}^{(X)}\left(\slash\!\!\!p \, , \, \varepsilon \right) \!\!&\simeq&\!\!
\frac{\alpha}{4\pi}
\left(-\slash\!\!\!p+4 m \right) \frac{\mu^{2\varepsilon}}{\varepsilon}
-\frac{\gamma \, \alpha}{4\pi} \left( - \slash\!\!\!p \, + 4 m \right)
\nonumber \\
&&
\hspace{-0.7cm}
- \, \frac{\alpha \, \chi^{\, 2}}{2\pi} \, \int_{0}^{1} dz \left[\left( 1-z \right) \, \slash\!\!\!p \, - 2 \, m \right] \,
\nonumber \\
&&
\hspace{-0.9cm}
\times\ln\left[ \frac{4\pi \mu^{2}}{ m_{X}^{\,2}(1-z)+m^{2}z-p^{\,2}z(1-z)} \right] \, ,
\hspace{1cm}
\end{eqnarray}
where $p^{2}<2 \, m^{2}+ 2 \, m_{X}^2$. The renormalized full propagator is represented by the expression
\begin{eqnarray}
s(\slash\!\!\!p)=\frac{Z_{2}}{\slash\!\!\!p-m-\Sigma(\slash\!\!\!p=m)} \, ,
\end{eqnarray}
where the $Z_{2}$-renormalization factor connects the electron bare field $\Psi_{0}$ to the physical field $\Psi$ via
$\Psi_{0}=\sqrt{Z_{2}} \, \Psi$. It is given by the on shell condition
\begin{eqnarray}
\left. Z_{2}=1+\frac{d \Sigma(\slash\!\!\!p)}{d \slash\!\!\!p}\right|_{\slash\!\!\!p=m} \; .
\end{eqnarray}
The electron physical mass is identified as being $m_{e}=m+\Sigma(\slash\!\!\!p=m)$, and the leading order
contribution to the electron's mass is
%
%
\begin{eqnarray}\label{SigmaQEDPodolskyepsilonp=m}
\Sigma_{1}\left( m \, , \, \varepsilon \right)
\!&\simeq&\! \frac{3m \alpha}{4\pi} \, \frac{\mu^{2\varepsilon}}{\varepsilon}
-\frac{3m\alpha \,\gamma}{4\pi}
\nonumber \\
&&
\hspace{-1cm}
+ \, \frac{3 m \alpha}{4\pi} \, \ln\left(\frac{4\pi\mu^{2}}{m^{2}} \right)
+\frac{5 m \alpha}{4\pi}
\nonumber \\
&&
\hspace{-1cm}
+ \, \frac{7m\alpha \, \chi^{\, 2}}{8\pi} - \frac{3m\alpha \, \chi^{\, 2}}{2\pi} \, \ln\left(\frac{m_{X}}{m} \right) \; ,
\end{eqnarray}
where we have assumed $m^{2}/m_{X}^{2} \ll 1$, and $1+\chi^{\, 2}\simeq 1$. This result gives the contribution
at the one loop approximation for the electron's renormalization mass. Thereby, the electron's physical mass has the finite
correction given by
\begin{equation}
\frac{m_{e}}{m}=1+\frac{5\alpha}{4\pi}
+ \frac{7\alpha \, \chi^{\, 2}}{8\pi}
- \frac{3\alpha \, \chi^{\, 2}}{2\pi} \ln\left(\frac{m_{X}}{m} \right) \; .
\end{equation}
Finally, the $Z_{2}$-factor is
\begin{eqnarray}
Z_{2}\!\!&=&\!\!1-\frac{\alpha}{2\pi} \frac{\mu^{2\varepsilon}}{\varepsilon}-\frac{\alpha \, \gamma}{2\pi}-\frac{\alpha}{\pi}
-\frac{\alpha}{4\pi} \ln\left(\frac{4\pi\mu^2}{m^2} \right)
\nonumber \\
&&
\hspace{-0.5cm}
-\frac{\alpha \, \chi^2}{4\pi} \, \ln\left(\frac{4\pi\mu^2}{m_{X}^2} \right)
-\frac{\alpha}{2\pi} \, \ln\left(\frac{m_{\gamma}}{m}\right) \; ,
\end{eqnarray}
Therefore, we have provided the renormalization result
for both the propagator and the electron-field with the contribution of the $X$-boson.


\section{The full $X$-propagator and the Uehling potential}

The analysis of the $X$-boson propagator is important in order to understand the $X$-boson physical mass.
We will see that the $m_{X}$-physical mass leads us to the calculation of the on shell complex renormalization.
Furthermore, its radiative correction contributes to the Yukawa potential obtained in section IV.
Let us start with the renormalized field of the $X$-boson which is defined by the relation
\begin{eqnarray}
X_{0}^{\, \mu}=\sqrt{Z_{X}} \, X^{\mu} \; ,
\end{eqnarray}
where the full renormalized propagator of the $X$-boson depends on the $Z_{X}$-factor, and it is given by the expression
\begin{eqnarray}\label{FullPropX}
\Delta_{\mu\nu}(k^{2})=-\frac{i \, \eta_{\mu\nu} \, Z_{X}^{-1}}{\left(1-\Pi\left(k^{2}\right) \right)k^{2}-m_{0X}^{\, 2}} \; .
\end{eqnarray}
The $\Pi(k^{2})$ is a scalar function that multiplies the transverse term in the vacuum polarization
\begin{eqnarray}
\Pi_{\mu\nu}(k^{2})=\Pi(k^{2})\left( \, \eta_{\mu\nu} k^{2}-k_{\mu} k_{\nu} \, \right) \; .
\end{eqnarray}
The conserved current guarantees that $k_{\mu}J^{\mu}=0$.    Hence, the terms like $k_{\mu}k_{\nu}/k^{2}$
are zero due to the term $J_{\mu} \, \Delta^{\mu\nu} \, J_{\nu}$, in the perturbation theory.
The on shell condition $k^{2}=m_{X}^{\, 2}$ fixes the propagator pole by using the $Z_{X}$-factor condition
\begin{eqnarray}
Z_{X}=\frac{1}{1-\Pi(m_{X}^{\, 2})} \; ,
\end{eqnarray}
thus, the $X$-propagator in Eq. (\ref{FullPropX}) is finite and it is given by
\begin{eqnarray}\label{FullPropXFinite}
\Delta_{\mu\nu}(k^{2})=-\frac{i \, \eta_{\mu\nu}}{\left[1-\Pi_{R}\left(k^{2}\right) \, \right]
k^{2}-m_{X}^{\, 2}} \; .
\end{eqnarray}
The scalar function $\Pi_{R}(k^2)=\Pi\left(k^{2}\right)-\Pi\left(m_{X}^{\, 2}\right)$ is finite, and
it cancels out the divergence that appears from the vacuum polarization.     Notice that for the physical mass $m_{X}$,
which is related to the unphysical mass $m_{0X}$ according to the relation $m_{0X}=\sqrt{Z_{X}^{-1}} \, m_{X}$. Thereby,
the renormalization factor for the $X$-boson mass is $Z_{m_{X}}=Z_{X}^{-1}$.
Using the previous rules, the expression of the vacuum polarization at one-loop approximation is given by the integral
\begin{equation}\label{Polvacuo1loop}
i \, \Pi_{1}^{\mu\nu}(m,k)=-\left(i \, e \, \chi\right)^2 \!
\int\frac{d^4p}{(2\pi)^4}
\,\mbox{tr}\left(\frac{i\gamma^{\mu}}{\slash\!\!\!p-m}\;\frac{i\gamma^{\nu}}{\slash\!\!\!p-\slash\!\!\!k-m} \right) ,
\end{equation}
which is the same expression of the standard QED considering the factor $\chi$.   Consequently,
the calculation of the vacuum polarization in this order is the same than the usual QED. Using the dimensional regularization, {\it i. e.},
$D=4 \rightarrow D=2\omega$, the previous integral can be calculated and the result is
\begin{eqnarray}\label{Pi2omega}
\Pi_{1}(m^{2},k^{2},\omega)=-\frac{\alpha \, \chi^{\, 2}}{\pi}\,(\mu^{2})^{2-\omega} \omega \, \Gamma(2-\omega) 
\nonumber \\
\times \int_{0}^{1} dx \, x \, (1-x)
\left[ \, \frac{4\pi\mu^{2}}{m^{2}-k^{2}x(1-x)} \, \right]^{2-\omega} ,
\hspace{0.5cm}
\end{eqnarray}
where $4m^{2}>k^{2}$. We isolate the divergent part by writing $\omega=2-\varepsilon$,
with $\varepsilon \rightarrow 0^{+}$ to obtain the result
\begin{eqnarray}\label{Pi2epsilon}
&&\Pi_{1}(m^{2},k^{2},\varepsilon)=-\frac{\alpha \, \chi^{\, 2}}{3\pi} \frac{\mu^{2\varepsilon}}{\varepsilon}
+\frac{\alpha \, \chi^{\, 2}}{6\pi}(2\gamma+1)
\nonumber \\
&&
-\frac{2 \alpha \, \chi^{\, 2}}{\pi} \! \int_{0}^{1} \! dx \, x \, (1-x)
\ln\left[ \, \frac{4\pi\mu^{2}}{m^{2}-k^{2}x(1-x)}\right] \, .
\hspace{0.5cm}
\end{eqnarray}
Using this result, the finite part that appears in Eq. (\ref{FullPropXFinite}) is given by the subtraction
\begin{equation}\label{PiR}
\Pi_{R}\left(k^{2}\right)
=\frac{2 \alpha \, \chi^{\,2}}{\pi}\int_{0}^{1} dx \,  x \, (1-x) \,
\ln\left[ \frac{m^{2}-k^{2} \, x \, (1-x)}{m^{2}-m_{X}^{2} \, x \, (1-x)} \right] \, .
\end{equation}
%
The integral in Eq. (\ref{PiR}) can be calculated for all values of $k^{2}$.   So, we obtain
%
%
%
%
%
\begin{eqnarray}\label{Pi2epsilon}
\Pi_{R}(k^{2})\!\!&=&\!\!-\frac{4\alpha \, \chi^{\, 2}}{3\pi} \left( \frac{m^2}{k^{2}} -\frac{m^{2}}{m_{X}^{2}} \right)
\nonumber \\
&&
\hspace{-1cm}
+\frac{\alpha \, \chi^{\, 2}}{3\pi}\left(1+\frac{2m^{2}}{k^{2}}\right) f(k^{2})
\nonumber \\
&&
\hspace{-1cm}
-\frac{\alpha \, \chi^{\, 2}}{3\pi}\left(1+\frac{2m^{2}}{m_{X}^{2}}\right) f(m_{X}^{2})
\; \; ,
\end{eqnarray}
where the function $f(k^{2})$ is defined by
\begin{equation}
f(k^{2})
=\left\{
\begin{array}{lll}
\!\!2 \sqrt{1-\frac{4m^{2}}{k^{2}}} \sinh^{-1}\!\! \left(\frac{\sqrt{-k^{2}}}{2m} \right)
\hspace{0.1cm} \mbox{if} \hspace{0.15cm} k^{2}<0 \; \; , \nonumber \\
\!\!\sqrt{\frac{4m^{2}}{k^{2}}-1} \cot^{-1}\!\! \left(\sqrt{\frac{4m^{2}}{k^{2}}-1}\right)
\hspace{0.1cm} \mbox{if} \hspace{0.15cm} 0<k^{2}\leq 4m^{2} , \nonumber \\
\!\!\sqrt{1-\frac{4m^{2}}{k^{2}}}
\left[ 2 \cosh^{-1}\!\! \left(\frac{\sqrt{k^{2}}}{2m} \right)-i\pi \right]
\hspace{0.1cm} \mbox{if} \hspace{0.15cm} k^{2}>4m^{2} \; .
\end{array}
\right.
\hspace{-1.5cm}
\nonumber \\
\end{equation}
In this expression, we can observe the appearance of an imaginary part, when $k^{2}>4m^{2}$.
This is the $X$-boson case where the on-shell condition $k^{2}=m_{X}^{2}$
fixes the inequality $m_{X}^{2}>4 m^{2}$, for the masses $m_{X}=17 \, \mbox{MeV}$ and $m=0.5 \, \mbox{MeV}$.
The imaginary part means the instability of the $X$-boson and, as a consequence, it decays into the virtual
electron-positron pair.

The result of the integral in Eq. (\ref{Pi2epsilon}) under the on-shell condition $k^{2}=m_{X}^{2}$ yields the $Z_{X}$-factor
\begin{eqnarray}
&&Z_{X} \simeq 1-\frac{\alpha \, \chi^{\, 2}}{3\pi} \frac{\mu^{2\varepsilon}}{\varepsilon}
+\frac{\alpha \, \chi^{\, 2}}{6\pi} \, \left(2\gamma+1\right)
\nonumber \\
&&
\hspace{-0.5cm}
-\frac{\alpha \, \chi^{\, 2}}{3\pi} \ln\left(\frac{4\pi \mu^{2}}{m^2}\right)\!
-i \, 2\alpha \, \chi^{\, 2}
+\frac{4\alpha \, \chi^{\, 2}}{\pi} \ln\left(\frac{m_{X}}{m}\right) \, .
\end{eqnarray}
Thereby, the $Z_{m_{X}}$-factor can be obtained, and the physical mass of the $X$-boson is
\begin{eqnarray}\label{mX/m}
&& \frac{m_{X}^{(R)}}{m_{0X}} \simeq 1-\frac{\alpha \, \chi^{\, 2}}{6\pi} \frac{\mu^{2\varepsilon}}{\varepsilon}
+\frac{\alpha \, \chi^{\, 2}}{12\pi} \, \left(2\gamma+1\right)
\nonumber \\
&&
\hspace{-0.5cm}
-\frac{\alpha \, \chi^{\, 2}}{6\pi} \ln\left(\frac{4\pi \mu^{2}}{m^2}\right)\!
-i \, \alpha \, \chi^{\, 2}
+\frac{2\alpha \, \chi^{\, 2}}{\pi} \, \ln\left(\frac{m_{X}}{m}\right) \, .
\end{eqnarray}
%

The general expression for the energy potential is given by
%
\begin{eqnarray}\label{UIntFourier}
U({\bf r})=- 4\pi\alpha \, \chi^{\, 2} \int \frac{d^3{\bf k}}{(2\pi)^{3}} \, e^{i\,{\bf k} \cdot {\bf r}} \, \Delta_{00}\left(-{\bf k}^{2}\right) .
\end{eqnarray}
Using the propagator in Eq.  (\ref{FullPropX}) into the energy potential in Eq. (\ref{UIntFourier}), where we have neglected the order $\alpha^{3}$ terms, after some algebraic manipulation, by using the approximation $m^2/m_{X}^{2}\ll 1$, the previous integral can be reduced to one quadrature in the following expression,
\begin{eqnarray}\label{UintdkSimpZeta}
&&U(r)=- \alpha \, \chi^{\, 2} \, \frac{e^{-m_{X} \, r}}{r}
-\frac{\alpha^2 \, \chi^{\, 4}}{3 \pi r} \, e^{-m_{X} \, r}
\nonumber \\
&&
-i \, \frac{\alpha^{2} \, \chi^{\,4}}{3 \, r} \left(1-\frac{m_{X}r}{2} \right) e^{-m_{X}r}
-\frac{2\alpha^{2} \chi^{\, 4}}{3\pi r} 
\nonumber \\
&&
\!\!\!\!\!\times\!\!\! \int_{1}^{\infty} \!\!\!\! d\xi \left(1+\frac{1}{2\xi^{2}} \right)
\!\frac{\left(\xi^{2}-1 \right)^{1/2}}{\xi^{2}}\!
\left(1-\frac{m^{2}}{4m_{X}^{2}\xi^{2}}\right)^{-2} \!\! e^{-2mr\xi} .
\nonumber \\
\end{eqnarray}
This $\xi$-integral can be called as the integral representation of the {\it Uehling potential} with the correction of the $X$-boson mass.
The $\xi$-integral is difficult to solve analytically, so we have to analyze it considering the asymptotic case $mr \gg 1$. For $mr \gg 1$, only the region $0 \leq \xi-1 \ll \left(mr\right)^{-1}$ contributes to the integral, so one can approximate $\xi \simeq 1$ 
to obtain the expression
\begin{eqnarray}\label{UintdkSimpZetaapprox}
U(r) &\!\!\simeq\!\!& - \, \frac{\alpha \, \chi^{\, 2}}{r} \, e^{-m_{X} \, r} \left(1+\frac{\alpha \, \chi^{2}}{3\pi} \right)
-\frac{\alpha^{2} \, \chi^{\, 4}}{4 \sqrt{\pi} r}
\, \frac{ e^{-2mr}}{\left(mr\right)^{3/2}}
\nonumber \\
&&
- \, i \, \frac{\alpha^{2} \,\chi^{\,4}}{3 \, r} \, \left(1-\frac{m_{X}r}{2} \right) e^{-m_{X}r}
\; ,
\end{eqnarray}
whenever $m\neq0$. Obviously, the imaginary part goes to zero when $m_{X} \rightarrow \infty$ (or $m_{X} \gg m$).
In this calculation, we do not take into account the loop corrections of neutrinos and quarks
(or pion and neutron).   However, it would be convenient to do it in the entire model involving the $SU_{L}(2)$-group,
concerning the unification of $X$-boson with the weak interaction.
%
%

\section{The corrections to the Vertex}

The vertex is another diagram of the model useful for the renormalization of the constant coupling.
Furthermore, it includes the electron's anomalous magnetic moment which is a famous  QED calculation
with an incredible agreement with the experimental result. 

We have two vertices in the $X$-boson model:
the usual one of QED, and the interaction of $X$ with the SM $\Psi$-fermions. As a consequence,
we have two $Z$-renormalization factors for each vertex diagram. Thus, the computation of the anomalous moment
motivates us to obtain a bound on the $\chi$-parameter.

Let us start with the QED photon-vertex where two diagrams contribute.   So we will denote it by the sum
\begin{eqnarray}
\Gamma^{\mu}\left(p \, ,\, p^{\prime}\right)=\Gamma_{1-QED}^{\mu}\!\left(p \, , \, p^{\prime}\right)
+\Gamma_{1-X}^{\mu}\!\left(p \, , \, p^{\prime}\right) \; ,
\end{eqnarray}
where the contribution of the X-boson propagator in this order is represented by the integral
\begin{eqnarray}\label{VertexQED1loop2omegaPart}
&&
\Gamma_{1-X}^{\mu}\left(p \, , \, p^{\prime}\right)=- \, e^{3} \, \chi^{\, 2} \, \int \frac{d^{4}k}{(2\pi)^{4}} \, 
\nonumber \\
&&
\times \, \frac{\gamma^{\alpha} \left( \slash\!\!\!k+\slash\!\!\!p^{\prime}+m \right) \gamma^{\mu} \left(\slash\!\!\!k+\slash\!\!\!p+m \right)
\gamma_{\alpha}}
{\left[ \left( k+p^{\prime} \right)^{\, 2}\!\!\!-\!m^{2} \right]\!\!
\left[ \left( k+p \, \right)^{2}\!\!\!-\!m^{2} \right] \!\! \left( k^{\, 2}\!-m_{X}^{\, 2} \right)} \; .
\end{eqnarray}
The QED photon-vertex has the similar expression exchanging
$e^{3} \, \chi^{\, 2} \rightarrow e^3$ and $m_{X} \rightarrow m_{\gamma}$ for the photon mass,
introduced to tame the infrared divergence.
This integral has a ultraviolet divergence by a simple power counting. Therefore, we will use the previous
technique of dimensional regularization to isolate the divergent term from the physical terms.
The divergent part has the result
\begin{eqnarray}
\Gamma^{\mu}\left(q^2\right)=- i \, e \, \mu^{2\epsilon} \gamma^{\mu} \,
\left[1+\frac{\alpha}{2\pi}\frac{\mu^{2\epsilon}}{\epsilon}+ \mbox{finite part} \right] \, .
\; \; \;
\end{eqnarray}
The vertex of $X$-boson with leptons has the correction at one loop given by
\begin{equation}
\Lambda^{\mu}\left(q^2\right)=- i \, e \, \chi \, \mu^{2\epsilon} \gamma^{\mu} \,
\left[1+\frac{\chi^2 \, \alpha}{2\pi}\frac{\mu^{2\epsilon}}{\epsilon}+ \mbox{finite part} \right] \, .
\; \; \;
\end{equation}
Thereby, we have obtained the necessary terms for the renormalization vertex.
The renormalization procedure will be performed in the next section.
Now, we are interested in the finite part of Eq. (\ref{VertexQED1loop2omegaPart}).

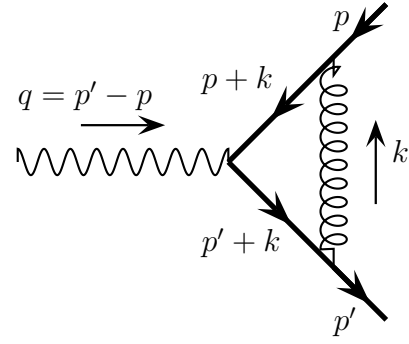
\begin{figure}[!h]
\begin{center}
\newpsobject{showgrid}{psgrid}{subgriddiv=1,griddots=10,gridlabels=6pt}
\begin{pspicture}(0,-0.7)(7,3.5)
\psset{arrowsize=0.2 2}
\psset{unit=0.7}
%
%
%
\pscoil[coilarm=0,coilaspect=0,coilwidth=0.5,coilheight=1.0,linecolor=black](1,2)(5,2)
\psline[linecolor=black,linewidth=0.7mm]{->}(8,5)(7.3,4.3)
\psline[linecolor=black,linewidth=0.7mm]{->}(8,5)(5.8,2.8)
\psline[linecolor=black,linewidth=0.7mm](6,3)(5,2)
\psline[linecolor=black,linewidth=0.7mm]{->}(5,2)(6.2,0.8)
\psline[linecolor=black,linewidth=0.7mm](6,1)(8,-1)
\psline[linecolor=black,linewidth=0.7mm]{->}(7,0)(7.7,-0.7)
\pscoil[coilarm=0.35,coilwidth=0.5,coilheight=1,linecolor=black](7,0)(7,4)
%
%
%
%
\psline[linecolor=black,linewidth=0.3mm]{->}(2.2,2.7)(3.8,2.7)
\put(1,3.1){\large$q=p^{\prime}-p$}
\psline[linecolor=black,linewidth=0.3mm]{->}(7.8,1.2)(7.8,2.8)
\put(8.1,2){\large$k$}
\put(7,4.6){\large$p$}
\put(7,-1.2){\large$p^{\prime}$}
\put(4.5,3.4){\large$p+k$}
\put(4.5,0.3){\large$p^{\prime}+k$}
%
%
%
\end{pspicture}
%
%
\caption{One-loop correction of the $X$-boson to the QED vertex diagram.}\label{VertexX}
\end{center}
\end{figure}

The finite part of the vertex that contains the form factors is consequently defined by the following difference given by
\begin{eqnarray}\label{GammaR}
\Gamma_{R}^{\mu}\left(q^2\right)=\Gamma^{\mu}\left(q^2\right)-\Gamma^{\mu}\left(q^2=0\right) \; .
\end{eqnarray}
%
The Gordon identity of the Dirac current can be used such that the finite part of Eq. 
(\ref{VertexQED1loop2omegaPart}) yields the relation
\begin{eqnarray}\label{baruGammauresult}
\Gamma_{R}^{\, \mu}\!\left(q^{\, 2} \right)= \gamma^{\mu} F_{1}\left(q^{2}\right)+i \, \frac{\sigma^{\mu\nu} q_{\nu}}{2m}
\, F_{2}\left(q^{2}\right) \, , \hspace{0.5cm}
\end{eqnarray}
where $F_{1}$ and $F_{2}$ are the form factors, and $q$
is defined as the photon's external momentum $q^{\mu}=p^{\prime \mu}-p^{\mu}$.
We have also used the usual on shell conditions for the fermion external momentum,
{\it i. e.}, $p^{\, 2}=p^{\prime \, 2}=m^{2}$. Thus, the form factors at one-loop
approximation are given by
\begin{eqnarray}\label{F1}
&&F_{1}\left( q^{2} \right)=\frac{\alpha}{2\pi}
\int_{0}^{1} dx \, dy \, dz \, \delta(x+y+z-1) 
\nonumber \\
&&
\times \, \frac{1+(2-z)^{2}+(1-x)(1-y) \, q^{2}/m^{2}}{(1-z)^2+z \, m_{\gamma}^{\,2}/m^{2}-x \, y \, q^{2}/m^{2}}
\nonumber \\
&&
+\frac{\alpha \, \chi^{\, 2}}{2\pi}
\int_{0}^{1} dx \, dy \, dz \, \delta(x+y+z-1) \, 
\nonumber \\
&&
\times \, \frac{1+(2-z)^{2}+(1-x)(1-y) \, q^{2}/m^{2}}{(1-z)^{2}+z \, m_{X}^{\,2}/m^{2}- x \, y \, q^{2}/m^{2} }
\; ,
\hspace{0.6cm}
\end{eqnarray}
and
\begin{eqnarray}\label{F2}
F_{2}\left( q^{2}\right)\!\!&=&\!\!
\frac{\alpha}{2\pi}
\int_{0}^{1} dx \, dy \, dz \, \delta(x+y+z-1)
\, 
\nonumber \\
&&
\times \, \,
\frac{2 z(1-z)}{(1-z)^2-x \, y \, q^{2}/m^{2}}
\nonumber \\
&&
\hspace{-0.5cm}
+\frac{\alpha \, \chi^{\, 2}}{2\pi}
\int_{0}^{1} dx \, dy \, dz \, \delta(x+y+z-1) \, 
\nonumber \\
&&
\hspace{-1cm}
\times \, \frac{2 z(1-z)}{ (1-z)^{2}+z \, m_{X}^{\,2}/m^{2}- x \, y \, q^{2}/m^{2} } \; .
\hspace{0.5cm}
\end{eqnarray}
The corresponding diagram is illustrated in the figure \ref{VertexX}.
%
%
%

The first factor $F_{1}$ is the origin of the infrared divergences in the model.
When $q^{\, 2}=0$, the photon mass $m_{\gamma}$ is the parameter that regularizes the infrared
divergence in the first integral of Eq. (\ref{F1}).
The second factor $F_{2}$ gives an important contribution to the electron
anomalous magnetic moment. This contribution appears when $q^{\, 2}=0$, where
the $X$-boson vertex carries the correction
\begin{equation}\label{F2=0}
F_{2}^{(e)}\left( 0 \right)=
\frac{\alpha}{2\pi}
+\frac{\alpha \, \chi^{\, 2}}{2\pi} \frac{m_{e}^{2}}{m_{X}^{2}}
\int_{0}^{1} \, dz \, \frac{2 \, z(1-z)^2}{z+(1-z)^{2} \, m_{e}^{2}/m_{X}^{2}}
\; .
\end{equation}
Here the first term is just the contribution of the ordinary QED.   The second integral
is the contribution of the mass $m_{X}$. We have expanded the previous integral for
$m^{2}/m_{X}^{2} \ll 1$.    Hence, we have obtained the result at the lower order
\begin{eqnarray}\label{F2=0}
F_{2}^{(e)} \, \left( \, 0 \, \right) \simeq
\frac{\alpha}{2\pi} \left( \, 1 + \frac{2}{3} \, \chi^{\, 2} \, \frac{m_{e}^{2}}{m_{X}^{\, 2}} \, \right) \; .
\end{eqnarray}
If we use the well known experimental uncertainty in the electron's anomalous magnetic moment, it leads us to the upper bound 
\begin{eqnarray}\label{inequality}
|\chi| \, \frac{m_{e}}{m_{X}} \lesssim 4.2 \times 10^{-5} \; .
\end{eqnarray}
%
%
%
Using the values $m_{e}=0.5 \, \mbox{MeV}$ and $m_{X}=17 \, \mbox{MeV}$, the upper bound for the $\chi$-parameter is
$|\chi| \lesssim 1.4 \times 10^{-3}$, which agrees with Eq. (\ref{chiconstraints}). This result just confirms
the  one present in the literature \cite{Marciano2014}. For the muon case, if we just consider its interaction with
$X$-boson, the $F_{2}$-form factor when $q^{2}=0$ is given by
\begin{eqnarray}\label{F2}
F_{2}^{(\mu)}\left(0\right)=
\frac{\alpha}{2\pi}
+\frac{\alpha \, \chi^{\, 2}}{2\pi}
\int_{0}^{1} dz \, \frac{2 z(1-z)^2}{ (1-z)^{2}+z \, m_{X}^{\,2}/m_{\mu}^{2}} \; ,
\end{eqnarray}
and using $m_{\mu}=105 \, \mbox{MeV}$, the $z$-integral has the result
\begin{eqnarray}\label{F2muon}
F_{2}^{(\mu)}(0)\simeq \frac{\alpha}{2\pi}\left(1+\frac{3}{2}\, \chi^2 \right) \, .
\end{eqnarray}
The experimental value for the muon $(g-2)_{\mu}$ factor is $(11659209 \pm 6) \times 10^{-10}$, so the result in Eq. 
(\ref{F2muon}) fixes the $\chi$-parameter as being $|\chi| \lesssim 2 \times 10^{-5}$.

%
%
\section{Renormalization}
\subsection{Renormalized perturbation theory}

To carry out the perturbative renormalization, both the renormalized fermion and gauge sectors are given by the Lagrangian
\begin{eqnarray}\label{renormalizedlagrangian}
&&\mathcal{L}=\bar{\Psi}\left( \, i \, \slash{\!\!\!\partial}-m \, \right)\Psi
+\bar{\Psi}\left( \, i \, \delta_{2} \, \slash{\!\!\!\partial}-\delta_{m} \, \right)\Psi
\nonumber \\
&&
-\frac{1}{4} \, F_{\mu\nu}^{2}-\frac{1}{2\alpha}\left(\partial_{\mu}A^{\mu}\right)^{2}
-\frac{\delta_{A}}{4} \, F_{\mu\nu}^{2}-\frac{\delta_{A}}{2\alpha}\left(\partial_{\mu}A^{\mu}\right)^{2}
\nonumber \\
&&
-\frac{1}{4} \, X_{\mu\nu}^{2}+\frac{1}{2} \, m_{X}^{2} \, X_{\mu}^{\, 2}
-\frac{1}{2\beta}\left(\partial_{\mu}X^{\mu}\right)^{2}
\nonumber \\
&&
-\frac{\delta_{X}}{4} \, X_{\mu\nu}^{2}-\frac{\delta_{X}}{2\beta}\left(\partial_{\mu}X^{\mu}\right)^{2}+\frac{1}{2} \, \delta_{m_{X}} \, X_{\mu}^{\, 2}
\nonumber \\
&&
-e \, \bar{\Psi} \, \, \slash{\!\!\!\!A} \, \Psi
-\delta_{e} \, \bar{\Psi} \, \, \slash{\!\!\!\!A} \, \Psi
-\chi \, e \, \bar{\Psi} \, \, \slash{\!\!\!\!X} \, \Psi
-\delta_{3} \, \bar{\Psi} \, \, \slash{\!\!\!\!X} \, \Psi \; ,
\; \; \; \;
\end{eqnarray}
where the relations between the bare and renormalized quantities are
\begin{eqnarray}
A^{(0)}_{\mu}=\sqrt{Z_{A}} \, A_{\mu}
\hspace{0.2cm} , \hspace{0.2cm}
\end{eqnarray}
and the counter-terms
\begin{eqnarray}\label{contratermosQED}
&&
Z_{A}=1+\delta_{A}
\hspace{0.1cm} , \hspace{0.1cm}
Z_{X}=1+\delta_{X}
\hspace{0.1cm} , \hspace{0.1cm}
\nonumber \\
&&
m+\delta_{m}=m_{0} \, Z_{2}
\hspace{0.1cm} , \hspace{0.1cm}
m_{0X}^{2} \, Z_{X}=m_{X}^{\, 2}+\delta_{m_{X}}
\hspace{0.1cm} , \hspace{0.1cm}
\nonumber \\
&&
e_{0} \, Z_{2} \, Z_{A}^{1/2}=e+\delta_{e}
\hspace{0.1cm} , \hspace{0.1cm}
e_{0} \, \chi \,  Z_{2} \, Z_{X}^{1/2}=\chi \, e+\delta_{3} \; .
\hspace{0.6cm}
\end{eqnarray}
We have introduced six counter-terms $\{ \delta_{A}, \delta_{X} , \delta_{m}, \delta_{m_{X}}, \delta_{e}, \delta_{3} \}$
to cancel out all the divergences that emerge from the Lagrangian in Eq. (\ref{renormalizedlagrangian}). Here,
all the parameters in the renormalized Lagrangian are finite, and we need six conditions to fix all these counter-terms.   They are
\begin{eqnarray}\label{CondRenor}
&&
\left. \Sigma(\slashed{p}) \right|_{\slashed{p}=m}=0
\hspace{0.1cm} , \hspace{0.1cm}
\left. \frac{d}{d\slashed{p}} \, \Sigma(\slashed{p}) \right|_{\slashed{p}=m}=0 \; ,
\nonumber \\
&&
\left. \Pi(k^{2})\right|_{k^{2}=0}=0
\hspace{0.1cm} , \hspace{0.1cm}
\left. \Pi_{X}(k^{2})\right|_{k^{2}= m_{X}^{2}}=0 \; ,
\nonumber \\
&&
\left. \Gamma^{\mu}(q^2) \right|_{q^2=0}=-\, i \, e \, \gamma^{\mu} \; ,
\nonumber \\
&&
\left. \Lambda^{\mu}(q^2) \right|_{q^2=m_{X}^{2}}=-\, i \, \chi \, e \, \gamma^{\mu} \; .
\hspace{0.8cm}
\end{eqnarray}
The first constraint fixes the physical electron mass and the second one fixes the renormalization of the fermion field.   The third one fixes the $Z_{A}$-factor and the fourth one fixes the $Z_{X}$-factor. The last two constraints fix both the vertex-photon, and $X$-boson-vertex, respectively. In sections V and VI,
we have obtained the $Z_{2}$- , $Z_{X}$- and $Z_{m_{X}}$-factors. Using the conditions in Eq. (\ref{CondRenor}), the counter-terms $\delta_{e}$
and $\delta_{3}$ can be determined such that we can obtain the relations $e_{0} \,  Z_{2} \, Z_{A}^{1/2}=e \, Z_{1}$ and
$e_{0} \,  Z_{2} \, Z_{X}^{1/2}=e \, Z_{3}$. The Ward identity guarantees that $Z_{1}=Z_{2}$, and the $Z_{A}$- and $Z_{3}$-factors
are given by
\begin{eqnarray}
Z_{A} \!\!&=&\!\! 1-\frac{\alpha}{3\pi}\frac{\mu^{2\epsilon}}{\epsilon}+ \mbox{finite part}
\\
Z_{3} \!\!&=&\!\! 1-\frac{\chi^2 \, \alpha}{2\pi}\frac{\mu^{2\epsilon}}{\epsilon}+ \mbox{finite part} \; .
\end{eqnarray}
Hence, we have determined all the $Z$-renormalization factors at one loop. This renormalization scheme
allows us to investigate the physical parameters as function of an arbitrary scale. In the next subsection, we will introduce
the renormalization group through the Callan-Symanzik equation. Its solution yields the running physical mass and the 
constant coupling.
%

%
%

%

\subsection{Renormalization group}

The Callan-Symanzik equation concerning the renormalization group for this model
is given by
\begin{eqnarray}\label{Callan}
&&
\left[ \, \mu \, \frac{\partial}{\partial \mu}+\beta(e) \, \frac{\partial}{\partial e}
+\beta_{\chi}(e_{\chi}) \, \frac{\partial}{\partial e_{\chi}}
\right.
\nonumber \\
&&
\left.
+m \, \gamma_{m}(e) \, \frac{\partial}{\partial m}
+m_{X} \, \gamma_{m_{X}}(e) \, \frac{\partial}{\partial m_{X}}
\right.
\nonumber \\
&&
\left.
-n \, \gamma_{A}(e)-n \, \gamma_{X}(e) \!\!\! \phantom{\frac{1}{2}}\right]\Gamma^{(n)}(e,e_{\chi},m,m_{X})=0
\; ,
\hspace{0.7cm}
\end{eqnarray}
where $\Gamma^{(n)}$ is the one-particle irreducible Green function of $n$-points, and $\mu$ is an arbitrary energy scale.
We will use in this work the notation $e_{\chi}= \chi \, e$, where the beta function $\beta_{\chi}$ is associated with the $X$-boson
vertex renormalization. The functions $ \beta $ , $\beta_{\chi}$ , $\gamma_{m}$ , $\gamma_{m_{X}} $ , $\gamma_{A}$ and $\gamma_{X}$
are related to the $Z$-renormalization factors by
\begin{eqnarray}\label{betagamma}
&&
\beta(e)=\mu \, \frac{\partial e}{\partial \mu}
\hspace{0.3cm} , \hspace{0.3cm}
\beta_{\chi}(e_{\chi})=\mu \, \frac{\partial e_{\chi}}{\partial \mu}
\nonumber \\
&&
\gamma_{m}(e)=-\frac{\mu}{2} \, \frac{\partial}{ \partial \mu}\ln Z_{m}
\hspace{0.1cm} , \hspace{0.1cm}
\gamma_{m_{X}}(e)=-\frac{\mu}{2} \, \frac{\partial}{ \partial \mu}\ln Z_{m_{X}}
\nonumber \\
&&
\gamma_{A}(e)=\frac{\mu}{2} \, \frac{\partial}{ \partial \mu}\ln Z_{A}
\hspace{0.1cm} , \hspace{0.1cm}
\gamma_{X}(e)=\frac{\mu}{2} \, \frac{\partial}{ \partial \mu}\ln Z_{X} \, .
\; \; \; \; \;
\end{eqnarray}
The function $\beta$ is kept constant in this $X$-boson framework. This occurs because, at high momenta, the mass $m_{X}=17 \, \mbox{MeV}$
is negligible, where $\beta$ is given explicitly by $\beta(e)=e^{3}/12\pi^{2}$, and the model is not asymptotically free.
Combining the functions in Eq. (\ref{betagamma}) and the relations in Eq. (\ref{contratermosQED}), it is easy to see that $\beta(e)=e\, \gamma_{A}(e)$, and consequently that $\gamma_{A}(e)=\alpha/3\pi$. The new functions $\beta_{\chi}$, $\gamma_{m_{X}}$ and $\gamma_{X}$ are 
\begin{eqnarray}
\beta_{\chi}(e) \!&=&\! \frac{e^3}{4\pi^2}\left(1-\frac{2 \,\chi^2}{3} \right) \; ,
\label{betachi}
\\
\gamma_{m_{X}}(e) \!&=&\! \gamma_{X}(e)=-\frac{e^{2} \, \chi^2}{12\pi^{2}} \; .
\label{gammaX}
\end{eqnarray}

The invariance of the Green function $\Gamma^{(n)}$ under scale transformation $\Gamma^{(n)}(e,e_{\chi},m,m_{X},\mu)=\Gamma^{(n)}(\bar{e}(t),\bar{e}_{\chi}(t),\bar{m}(t),\bar{m}_{X}(t),\bar{\mu}(t)=\mu\,e^{t})$ leads us to the
{\it effective coupling constants} $\bar{e}(t)$, $\bar{e}_{\chi}(t)$ and effective masses $\bar{m}(t)$ and $\bar{m}_{X}(t)$ as function
of the dimensionless scale $t$, all of them satisfying the equations
\begin{eqnarray}
\frac{\partial \bar{e}}{\partial t} \!\! &=& \!\! \beta(\bar{e}) \; \; \; ,
\label{eeffective}
\\
\frac{\partial \bar{e}_{\chi}}{\partial t} \!\! &=& \!\! \beta(\bar{e}_{\chi}) \; \; \; ,
\label{echieffective}
\\
\frac{\partial \bar{m}}{\partial t} \!\! &=& \!\! \bar{m}(t) \, \gamma_{{m}}(\bar{e}(t)) \; ,
\label{meffective}
\\
\frac{\partial \bar{m}_{X}}{\partial t} \!\! &=& \!\! \bar{m}_{X}(t) \, \gamma_{{m}_{X}}(\bar{e}(t)) \; ,
\label{mXeffective}
\end{eqnarray}
where $\bar{e}(t=0)=e$, $\bar{e_{\chi}}(t=0)= \chi \, e$, $\bar{m}(t=0)=m$ and $\bar{m}_{X}(t=0)=m_{X}=17 \, \mbox{MeV}$.
The solution of Eqs. (\ref{eeffective}) and (\ref{meffective}) provide the known results of QED for running coupling constant
and running electron mass. The solutions of (\ref{echieffective}) yields the running $X$-boson vertex with fermions
\begin{eqnarray}\label{et}
\bar{e}_{\chi}(t)= \chi \, e \, \left(1-\frac{\chi^2 \, e^{2} \, t}{2\pi^{2}}\right)^{-1/2} \; .
\end{eqnarray}
The previous function imposes the vertical asymptote at $t=2\pi^2/\left(\chi^2\, e^2\right)$.
This corresponds to the so called {\it Landau singularity}. The $\bar{e}_{\chi}$-running function is plotted in figure 5.
\begin{figure}[h]
\centering
\includegraphics[scale=0.45]{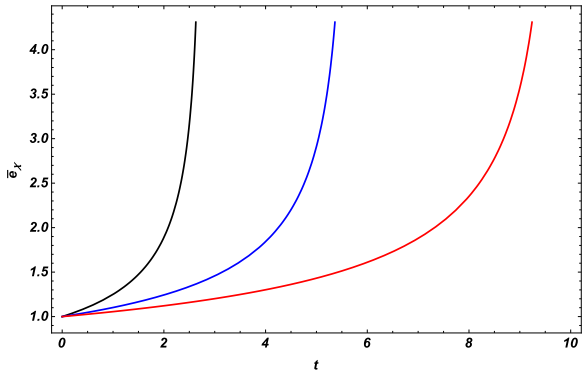}
\caption{The running $\bar{e}_{\chi}/(\chi \, e)$-coupling constant as function of an arbitrary dimensionless $t$-scale, for some $\chi$-values.}
\label{Constet}
\end{figure}
\noindent
Solving Eq. (\ref{mXeffective}), we obtain that
\begin{eqnarray}\label{LWMassEffective}
\bar{m}_{X}(t)= 17 \, \mbox{MeV} \, \left(1-\frac{\chi^2 \, e^{2} \, t}{2\pi^{2}}\right) \; ,
\end{eqnarray}
which yields the $X$-boson running mass as a function of arbitrary $t$-scale.
%


\section{Conclusions and Final Remarks}

Recently we were aware of the introduction of a concept that affirms that a new neutral boson explains the experimental anomalies that emerge from the $8$-Beryllium nuclear
decay $8 \, Be^{\,\ast} \rightarrow 8 \, Be + X$.
The solution of this puzzle implies that the invariant mass of the $X$-boson must be around $m_{X}=17 \, \mbox{MeV}$.
This conjecture plays a fundamental role in a possible new physics at the $\mbox{MeV}$-scale, so that it could be the announcement of a fifth fundamental interaction. Furthermore, the $X$-boson couples kinetically through the $\chi$-mixing parameter with the usual massless photon. Other important property is the protophobic interaction of $X$-boson with the nucleons of the SM. Thereby, the $X$-boson introduces an extra Abelian group $U'(1)$ in the unification of the fundamental interactions.

With these ideas in mind, in this work we have investigated the $U(1) \times U'(1)$ model with kinetic mixing in the gauge sector, which can describe
the interaction between the new $X$-boson and the leptons of the SM. The Higgs model was introduced to give the mass $m_{X}=17 \, \mbox{MeV}$,
that consequently fixes a VEV-scale around $v=6.3 \, \mbox{GeV}$, by the recent experimental constraints. Thus,
the hidden Higgs is estimated to have a mass within the range of GeV-scale, with a lightest mass in relation to the usual Higgs
of $125 \, \mbox{GeV}$ of the SM. After the spontaneous symmetry breaking mechanism, we have a renormalizable and unitary model in the $R_{\xi}$-gauge,
with an ending electromagnetic $U(1)_{em}$ symmetry.

After that, we have discussed the interaction between the $X$-boson and the leptons through
the elements of QFT, like the decay rate, where we have calculated a $X$-lifetime in the range of
$8.3 \times 10^{-15} \, \mbox{s} < \tau < 4.2 \times 10^{-12} \, \mbox{s}$, just taking into account the decays:
$X \rightarrow \, e^{+} \, e^{-}$, and in neutrino's case, $X \rightarrow \, \bar{\nu} \, \nu$. As a second application,
the simplest case is
the electron-positron scattering into muon-anti-muon pair, {\it i. e.},
$e^{+} \, e^{-} \, \stackrel{X}{\longrightarrow} \, \mu^{+} \, \mu^{-}$. 

Following the usual QFT, we have computed the contributions of the $X$-boson mass to the electron physical mass. The perturbation
theory for the $X$-boson full propagator was obtained.    We have seen that the vacuum polarization at one-loop gave a
contribution to the Yukawa potential. The on-shell renormalization indicates the appearance of a complex
contribution, as for example, in the case of the $X$-boson physical mass in Eq. (\ref{mX/m}), and in the
Uehling potential calculation. In fact, this complex renormalization scenario is a consequence of the $X$-boson instability,
that decays into the $e^{+} \, e^{-}$-pair. The correction to the QED vertex was calculated, and
the electron's anomalous magnetic moment $g-2$ estimates the $\chi$-parameter around $|\chi| \lesssim 1.4 \times 10^{-3}$, which is
in agreement with the literature results. The muon anomalous magnetic moment also was obtained with the $X$-boson correction, which uncertainty of $(g-2)_{\mu}$ fixes the $\chi$-parameter in $|\chi| \lesssim 2 \times 10^{-5}$.

We have introduced the renormalized model and the renormalization conditions to fix the
physical parameters. Thereby, all the renormalization factors were obtained: for the physical fields, masses and the coupling constants.
After that, we have applied these results in a renormalization group scheme to obtain the behavior of both, the current $X$-mass and the $X$-boson
coupling constant with leptons.

It is clear that we have just considered a part of a bigger
model in order to include neutrinos and quarks. The extended model $SU_{L}(2) \times U_{R}(1) \times U'(1)$ is the next candidate to include the 
corrections to neutrinos/quarks to the $X$-propagator, together with the contribution of the $X$-propagator to the $(g-2)_{\mu}$ muon factor.
Furthermore, there is also a perspective to include a hidden fermion sector that could have a dark matter feature which interacts with the $X$-Boson.

Other phenomenological approach would be to investigate the $SU_{L}(2) \times U_{R}(1) \times U'(1)$ symmetry group with a dark photon $A'$ of mass bounded by $m_{A'} \, \lesssim
\, 8 \, \mbox{GeV}$, and to analyze its interactions with the fermion part of the SM and a possible content of a dark matter fermion. 
It is a ongoing research that will be published elsewhere.
%
%


%

\section*{Acknowledgments}

\ni The authors thank professor Jos\'e Abdalla Hela\"yel-Neto for valuable discussions.
E.M.C.A.  thanks CNPq (Conselho Nacional de Desenvolvimento Cient\' ifico e Tecnol\'ogico), Brazilian scientific support federal agency, for partial financial support, Grant number 302155/2015-5 and the hospitality of Theoretical Physics Department at Federal University of Rio de Janeiro (UFRJ), where part of this work was carried out.



\end{document}